\date{}
\documentclass[aps,pra,showpacs]{revtex4-1}
\usepackage{graphicx,psfrag,amsmath,amssymb,amsfonts,latexsym,color,dcolumn}
\begin{document}
\title{Derivative expansion for the Casimir effect at zero and finite
temperature in $d+1$ dimensions}
\author{C\'esar D. Fosco$^{1,2}$}
\author{Fernando C. Lombardo$^3$}
\author{Francisco D. Mazzitelli$^{1}$}

\affiliation{$^1$ Centro At\'omico Bariloche,
Comisi\'on Nacional de Energ\'\i a At\'omica,
R8402AGP Bariloche, Argentina}
\affiliation{$^2$ Instituto Balseiro,
Universidad Nacional de Cuyo,
R8402AGP Bariloche, Argentina}
\affiliation{$^3$ Departamento de F\'\i sica {\it Juan Jos\'e
 Giambiagi}, FCEyN UBA, Facultad de Ciencias Exactas y Naturales,
 Ciudad Universitaria, Pabell\' on I, 1428 Buenos Aires, Argentina - IFIBA}
\date{today}
\begin{abstract} 
We apply the derivative expansion approach to the Casimir effect for a
real scalar field in $d$ spatial dimensions, to
calculate the next to leading order  term in that expansion, namely,
the first correction to the proximity force approximation.  The field
satisfies either Dirichlet or Neumann boundary conditions on two static
mirrors, one of them flat and the other gently curved.  We show
that, for Dirichlet boundary conditions, the next to leading order term in the Casimir
energy is of quadratic order in derivatives, regardless of the number of
dimensions. Therefore it is local, and determined by a single coefficient.
We show that the same holds true, if $d \neq 2$, for a  field which
satisfies Neumann conditions. When $d=2$, the next to leading order  term becomes nonlocal in
coordinate space, a manifestation of the existence of a gapless excitation
(which do exist also for $d> 2$, but produce sub-leading terms).  
 
We also consider a derivative expansion approach including thermal fluctuations
of the scalar field.  We
show that, for Dirichlet mirrors, the next to leading order term in the free energy is also
local for any temperature $T$. Besides, it interpolates
between the proper limits:  when  $T \to 0$ it tends to the one we
had calculated for the Casimir energy in $d$ dimensions, while for $T \to \infty$ it
corresponds to the one for a theory in $d-1$ dimensions, because of the expected dimensional
reduction at high temperatures.  
For Neumann mirrors in $d=3$, we find a nonlocal next to leading order  term for any $T>0$.
\end{abstract}

\maketitle
\section{Introduction}\label{sec:intro}
The determination of the Casimir force \cite{books} for a quite general situation,
namely, when the geometry of the problem is characterized by two rather
arbitrary surfaces, is interesting and potentially useful.  Those surfaces
may correspond, for example, to the boundaries of two mirrors.
Alternatively, the surfaces themselves may describe zero-width (`thin')
mirrors, which will be the situation considered in this paper. Yet another possibility is that those surfaces may be the
interfaces between media with different electromagnetic properties,
occupying different spatial regions.  In situations like the ones above, it
may be convenient to think of the Casimir energy as a {\em functional\/} of
the functions determining the surfaces.  Of course, it is generally quite
difficult to compute that functional for arbitrary surfaces. Rather, exact
results are only available for highly symmetric configurations, the
simplest of which being the case of two flat, infinite, parallel plates.
Taking advantage of the simplicity of the result for this highly symmetric
configuration, the proximity force approximation (PFA) \cite{Derjaguin, PFA} provides an accurate
method to calculate the Casimir energy when the surfaces are gently curved,
almost parallel, and close to each other. Introduced by Derjaguin many
years ago~\cite{Derjaguin} to compute Van der Waals forces, this
approximation consists of replacing both surfaces by a set of parallel
plates. The energy is then calculated as the sum of the Casimir energies
due to each pair of plates, each plate paired only with the nearest one in
the other mirror. 

In a recent work~\cite{pfa_nos}, we have shown that the PFA can
be put into the context of a derivative expansion (DE) for the Casimir
energy, when the latter is regarded as a functional of the functions that
define the shapes of the mirrors. Indeed, the leading order term in
this expansion, which contains no derivatives, does reproduce the PFA,
while the higher order ones account for the corrections.  
In that article we considered, for the sake of simplicity, a massless
quantum scalar field satisfying Dirichlet boundary conditions on two
surfaces. One of them was assumed to be flat, and such that if coordinates
were chosen so that $x_3=0$, the other surface could be described by a
single function: $x_3=\psi(x_1,x_2)$. 

Since the form of the possible terms in the DE may be determined by dimensional
analysis plus symmetry considerations, what is left is the calculation of
their respective coefficients. Note that those coefficients are
`universal', in the sense that they are independent of the shapes of the
mirrors (at least for smooth surfaces).  Therefore one can fix those coefficients
completely, from the knowledge of their values for a particular surface, or
for a family of surfaces.  We used, to that effect, a particular family of
surfaces, namely, those obtained by an expansion up to the second order in 
$\eta$, which is the (assumed small) departure from flat parallel mirrors: 
\mbox{$\psi = a + \eta$}, \mbox{$\eta << a$}. 
The coefficients determined from this family were then used to fix the
coefficients of the first two terms in the DE, which can then be used to
calculate the Casimir energy for more general (but smooth) surfaces.

This approach has been generalized by Bimonte et al~\cite{pfa_mit1,bimonteT} in many
directions. For example, to the case of two curved, perfectly conducting
surfaces, for scalar fields satisfying Dirichlet or Neumann boundary conditions, 
and also to the electromagnetic case and imperfect boundary conditions.  
As a validity check, it has been shown that,
whenever analytic results are available for particular geometries, the
corresponding DE does reproduce both the PFA and its next to leading order (NTLO)
correction~\cite{pfa_nos,pfa_mit1}.  The DE approach has also been applied successfully
to compute the electrostatic interaction between perfect conductors \cite{nos_annphys}.  

In~\cite{pfa2_nos}, we have extended our previous work~\cite{pfa_nos} to
the case of the electromagnetic field coupled to two thin, imperfect
mirrors, described by means of the vacuum polarization tensors localized on
the mirrors. We have also calculated the NTLO to the PFA static Casimir
force. For the particular case of mirrors described by a single
dimensionless quantity, we have computed the leading and NTLO
corrections as a function of that quantity. We found that the 
absolute value of the NTLO correction falls
down rather quickly for imperfect mirrors \cite{pfa2_nos}.  

In this article, we apply the DE approach, to models
where the fluctuating field is coupled to two perfect mirrors, 
$L$ and $R$, at zero or finite temperature, in $d+1$ spacetime
dimensions.  Special consideration shall be given to the limiting cases
where the fluctuations are either thermal or quantum mechanical, i.e,
infinite or zero temperature. 

The role of the fluctuating field is played by a massless real scalar field
$\varphi$, with either Dirichlet or Neumann boundary conditions on both
mirrors.  This analysis is of interest because of several reasons: on the
one hand, as a first step towards incorporating thermal effects into the
DE (a fuller treatment should also include finite conductivity corrections
along with finite temperature corrections \cite{bimonteT}). On the other hand, as we will
see, this more general analysis will lead us to a clearer  physical picture
of the validity of the DE and will also shed some light
about possible extensions and improvements.  
 
We will mostly consider  the first two terms in the DE; in order to fix their coefficients,
we follow the procedure of expanding the vacuum energy up to the second
order in $\eta$ and then extracting the coefficients from the corresponding
momentum space kernel.  Thus, in what may be considered as a byproduct of
our approach, we also present the general result for that kernel, 
valid for any $d$,  both for the Dirichlet and Neumann cases. 

This kernel for the quadratic term in $\eta$ (regarded now as a field) may
be interpreted as a contribution to its 2-point one-particle irreducible
function, due to a one-loop $\varphi$ contribution, which fluctuates
satisfying the proper boundary conditions.
The coefficients of the DE up to the NTLO result from that kernel, from its expansion up to 
the second order in $k$, the $\eta$-field momentum, as one would do in an
effective field theory approach~\cite{Donoghue}.

When that expansion to  order $k^2$ does exist, the NTLO terms are of
second order in derivatives, and therefore spatially local. 
However, as in any quantum correction to a $2$-point function, we know that
non-analytic contributions may arise when the external momentum reaches
the threshold to excite modes of the field in the loop. In the present
case, since we expand around $k=0$, those non-analyticities may
only originate in the existence of massless modes. When present, they result in
contributions which, being nonlocal in space, are similar to the ones that
appear in the context of effective field theories, when including the
effect of massless virtual particles~\cite{Donoghue}. 
We shall see that this kind of non-analyticity does indeed appear, for Neumann
conditions, in the form of branch cuts. The physical reason being that when
both mirrors impose Neumann conditions there are transverse gapless modes for the
fluctuating field. However, except for $d=2$ and zero temperature ($T=0$), or $d=3$
and $T>0$, those non-analyticities are of higher order than the NTLO, when
$k \to 0$.  
In this, the only `pathological' case ($d=2$ and $T=0$ or $d=3$ and $T>0$), the
real time version of the kernel has, if rotated to real time, a logarithmic 
branch cut at zero momentum of the form $k^2 \log(k^2)$, which overcomes
the $k^2$ term (which is also present).

We also study the NTLO term as a
function of temperature. In the particular case of $d=3$, we shall see that for Dirichlet boundary
conditions it depends smoothly on the temperature. For Neumann boundary
conditions, as an indirect consequence of the non-analyticity at $d=2$ and
zero temperature, the Neumann NTLO term is also non-analytic at any non zero
temperature. 

This paper is organized as follows. In the next Section we introduce the
system and summarize the approach we follow to calculate the free energy
$\Gamma_\beta$.  In Section~\ref{sec:Dirichlet} we discuss  the DE for Dirichlet
boundary conditions on thin, perfect mirrors in $d+1$ dimensions,
discussing the zero and high (infinite) temperature limits. We apply those
results to evaluate the Casimir interaction energy between a sphere and a
plane at very high temperature.  Section~\ref{sec:Neumann} is devoted to
study the DE at zero and high temperatures limits, for a real scalar field
with Neumann boundary conditions on the mirrors. The special cases of $d=2$ with $T=0$,
and $d=3$ with $T>0$,
are singled out and dealt with in subsection~\ref{ssec:nonanalit}.
Higher order terms in the DE are analyzed in Section~\ref{sec:hoDE}. Finally,
in Section~\ref{sec:conclusions}, we summarize our conclusions. The Appendices contain
some details of the calculations.
\section{The system}\label{sec:thesys}
We shall adopt Euclidean conventions, whereby the spacetime metric is the
identity matrix, and spacetime coordinates are denoted by $x^\mu = x_\mu$
\mbox{($\mu=0,1,\ldots,d$)}, $x_0$ being the imaginary time and $x_i$,
($i=1,\ldots,d$) the spatial Cartesian coordinates.

Regarding the geometry of the system, we shall assume that one of the surfaces, 
$L$, is a plane, while the other, $R$, is such that it can be described by a single Monge
patch:
\begin{equation}\label{eq:deflr}
L) \;\; x_d \,=\, 0 \,\;\;\;\;\;\; R) \;\; x_d \,=\,
\psi(x_0,x_1,\ldots,x_{d-1}) \;.
\end{equation}

We have included for $R$ in (\ref{eq:deflr}) a more general, time-dependent
boundary, in spite of the fact that we are interested in the {\em
static\/} Casimir effect (SCE).  We shall, indeed, at the end of the
calculations, impose the condition that the boundaries are time-independent: 
$\psi = \psi(x_1,\ldots,x_{d-1})$, 
but it turns out to be convenient to keep the
more general kind of boundary condition at intermediate stages of the
calculation.  In this way, the treatment becomes more symmetric, and one
may take advantage of that to simplify the calculation. Besides,
although it is not our object in this paper, one could rotate back some of
the results thus obtained for a non-static $\psi$ to real time, in order to consider
a dynamical Casimir effect (DCE) situation. 

We follow a functional approach to calculate the free energy
$\Gamma_\beta(\psi)$, or its zero temperature limit $E_{\rm vac}(\psi)$, the vacuum
energy. Both are functionals of $\psi$, that defines the shape
of the $R$ mirror (the plane mirror $L$ is assumed to be fixed at $x_d=0$). 
$\Gamma_\beta$ is also a function of the inverse temperature $\beta \equiv T^{-1}$ (we use units
such that Boltzmann constant $k_B=1$).

In the functional approach, which we shall follow, both objects are
obtained by performing a functional integration; indeed:
\begin{equation}\label{eq:defgbpsi}
	\Gamma_\beta(\psi)\,=\,-\frac{1}{\beta}\,\log\big[\frac{{\mathcal
	Z}_\beta(\psi)}{{\mathcal Z}_\beta^{(0)}}\big]\;.
\end{equation}   
where ${\mathcal Z}_\beta(\psi)$ is the partition function; it may be
obtained by integrating over field configurations that satisfy the
corresponding boundary conditions at $L$ and $R$, and are also periodic
(with period $\beta$) in the imaginary time coordinate $x_0$ (Matsubara
formalism).  ${\mathcal Z}_\beta^{(0)}$ denotes the partition function in
the absence of the mirrors; therefore it corresponds to a relativistic 
free Bose gas.

$E_{\rm vac}(\psi)$ is then obtained by taking the limit:
\begin{equation}
E_{\rm vac}(\psi)\,=\,\lim_{\beta \to \infty} \Gamma_\beta(\psi)\equiv \Gamma_\infty(\psi)\;.
\end{equation} 

It is worth noting that in (\ref{eq:defgbpsi}) $\psi$ has to be
time-independent for $\Gamma_\beta(\psi)$ to be a free energy. We do
however calculate objects like ${\mathcal Z}_\beta(\psi)$ for
configurations that may have a time dependence, keeping the same
notation.

To avoid an unnecessary repetition of rather similar expressions, we shall
write most of the derivations within the context of a finite temperature
system, presenting their zero temperature counterparts at the end of the
calculations.

We shall consider a real scalar field, with either Dirichlet or Neumann
boundary conditions.  In both cases, the general setup has a similar
structure, but there are also some important differences. Mostly, they come from
the different infrared behaviour of their respective Green's functions, and
the impact that that behaviour has on the DE. 
Accordingly, we present them in two separate sections.
\section{Dirichlet boundary conditions}\label{sec:Dirichlet}
We start from the functional representation of ${\mathcal Z}_\beta(\psi)$:
\begin{equation}\label{eq:defzetabeta}
	{\mathcal Z}_\beta(\psi)\;=\; \int {\mathcal D}\varphi \;
	\delta_L(\varphi) \; \delta_R(\varphi) \; e^{-{\mathcal
	S}_0(\varphi)} \;,
\end{equation}
where $\delta_A(\varphi)$, $A=L,R$, is a functional $\delta$ function which imposes
Dirichlet boundary conditions on the respective mirror, while ${\mathcal
S}_0$ is the free Euclidean action for a massless real scalar field in
$d+1$ dimensions, at finite temperature:
\begin{equation}\label{eq:defs0}
	{\mathcal S}_0 \,=\, \frac{1}{2} \int_0^\beta dx_0 \int d^dx \,
	(\partial \varphi)^2 \;,
\end{equation}
with periodic conditions for $\varphi$ in the time-like coordinate, namely,
$\varphi(x_0,{\mathbf x}) = \varphi(x_0 +\beta,{\mathbf x})$, for all
${\mathbf x}\in {\mathbb R}^{(d)}$. 

To proceed, one should then exponentiate the $\delta$-functionals by
introducing two auxiliary fields, $\lambda_L$ and $\lambda_R$, functions of
$x_\parallel \equiv (x_0,x_1,\ldots,x_{d-1}) \equiv (x_0,{\mathbf x}_\parallel)$, 
also satisfying periodic boundary conditions in the $x_0$ coordinate. 
In the Dirichlet case, we have:
\begin{eqnarray}\label{eq:deltexp}
	\delta_L(\varphi)  &=& \int {\mathcal D}\lambda_L \; e^{i \int
	d^dx_\parallel \, \lambda_L(x_\parallel) \, \varphi(x_\parallel,0)} \nonumber\\
	\delta_R(\varphi)  &=& \int {\mathcal D}\lambda_R \; e^{i \int
	d^dx_\parallel \, \sqrt{g (x_\parallel)} \, \lambda_R(x_\parallel) \,
	\varphi(x_\parallel,\psi(x_\parallel))} \;, 
\end{eqnarray}
where $g$ is the determinant of $g_{\alpha\beta}$, the induced metric on $R$:
\begin{equation}
	g_{\alpha\beta}(x_\parallel) \,=\, \delta_{\alpha\beta} +
	\partial_\alpha \psi(x_\parallel) \partial_\beta
	\psi(x_\parallel) \;,
\end{equation}
\begin{equation}
\Rightarrow \; g(x_\parallel) \,=\, 1 + (\partial\psi(x_\parallel))^2 \;.
\end{equation}
We have adopted the convention that indices from the beginning of the Greek
alphabet run from $0$ to $d-1$.

Using the exponential representations above in (\ref{eq:defzetabeta}), one
derives the alternative expression:
\begin{equation}\label{eq:zetabeta1}
	{\mathcal Z}_\beta(\psi)\;=\; \int {\mathcal D}\varphi \, {\mathcal
	D}\lambda_L {\mathcal D}\lambda_R \; e^{-{\mathcal S}_0(\varphi)
	\,+\, i \int d^{d+1}x J_D(x) \varphi(x)} \;,
\end{equation}
where the `Dirichlet current' $J_D(x)$ is given by:
\begin{equation}
	J_D(x) \;=\; \lambda_L(x_\parallel) \, \delta(x_3) \,+\, 
	\lambda_R(x_\parallel) \,\sqrt{g(x_\parallel)} 
 \delta(x_3 - \psi(x_\parallel))  \;.
\end{equation}
It is possible to get rid of the $\sqrt{g(x_\parallel)}$ factor above just by 
redefining  $\lambda_R$: $\lambda_R(x_\parallel) \to
\lambda_R(x_\parallel)/\sqrt{g(x_\parallel)}$. This redefinition induces a
nontrivial Jacobian. However, this Jacobian is independent of the distance
between the mirrors,  therefore irrelevant to the calculation of their
relative Casimir force; hence we discard it.

The integral over $\varphi$, a  Gaussian, yields:
\begin{equation}\label{eq:zetabeta2}
	{\mathcal Z}_\beta(\psi)\;=\; {\mathcal Z}_\beta^{(0)} \,
	\int {\mathcal D}\lambda_L {\mathcal D}\lambda_R \;
	e^{- \frac{1}{2} \int_{x_\parallel,x'_\parallel}
		\lambda_A(x_\parallel) {\mathbb T}_{AB}(x_\parallel,x'_\parallel)
	\lambda_B(x'_\parallel) }, 
\end{equation}
where we have introduced the objects:
\begin{align}
{\mathbb T}_{LL}(x_\parallel,x'_\parallel) &=\langle
x_\parallel,0|(-\partial^2)^{-1} |x'_\parallel,0\rangle \\
{\mathbb T}_{LR}(x_\parallel,x'_\parallel) &=\langle
x_\parallel,0|(-\partial^2)^{-1}
|x'_\parallel,\psi(x'_\parallel)\rangle \\
{\mathbb T}_{RL}(x_\parallel,x'_\parallel) &=\langle
x_\parallel,\psi(x_\parallel)|(-\partial^2)^{-1} 
|x'_\parallel,0\rangle \\
{\mathbb T}_{RR}(x_\parallel,x'_\parallel) &=\langle
x_\parallel,\psi(x_\parallel)|(-\partial^2)^{-1} 
|x'_\parallel,\psi(x'_\parallel)\rangle 
\end{align}
where we use a ``bra-ket'' notation to denote matrix elements of operators,
and
\begin{equation}
	\langle x|(-\partial^2)^{-1} |y\rangle \,=\, \sum_{n=-\infty}^{\infty}\int
	\frac{d^dk}{(2\pi)^d} \, \frac{e^{i(\omega_n (x_0-y_0)+ \mathbf {k} \cdot (\mathbf{x}-\mathbf{y}))}}{(\omega_n^2+\mathbf{k}^2)}\,\equiv\,
	\Delta (x-y)\;,
\end{equation}
where we have introduced the Matsubara
frequencies: $\omega_n \equiv \frac{2\pi n}{\beta}$, $n \in {\mathbb Z}$.
The free energy $\Gamma_\beta(\psi)$ is then
\begin{equation}\label{eq:fbd}
	\Gamma_\beta(\psi) \;=\; \frac{1}{2 \beta} {\rm Tr} \log {\mathbb T} \;.
\end{equation}
where $\psi$ is regarded as time independent, something that one can impose
at the end of the calculation.

$\Gamma_\beta$ still contains `self-energy' contributions, i.e., contributions
invariant under the rigid displacement $\psi(x_\parallel) \to
\psi(x_\parallel) +\epsilon$. Since we are just interested in the Casimir force
we shall neglect them altogether whenever they emerge in the calculations below.

\subsection{Derivative expansion}\label{ssec:dderivative}
The DE is implemented by following the same idea and
approach introduced in~\cite{pfa_nos}. The calculation is, in many
aspects, identical to the one in~\cite{pfa_nos}, with the only
differences in the number of dimensions and in the fact that the time
coordinate is compact (periodic), so frequency integrations have to be replaced
by sum over Matsubara frequencies.  Therefore, we do not repeat all the steps
presented there, rather, we limit ourselves to convey the relevant results. 

First, we note that in the DE approach applied to this case, keeping up to two
derivatives, the Casimir free energy can be written as follows:
\begin{equation}
	\Gamma_\beta (\psi) \;=\; \int d^{d-1}{\mathbf x}_\parallel \, 
\Big\{ b_0(\frac{\psi}{\beta}) \frac{1}{[\psi({\mathbf x}_\parallel)]^d} 
\,+\, 
b_2(\frac{\psi}{\beta}) \, \frac{(\partial\psi)^2}{[\psi({\mathbf x}_\parallel)]^d} 
\Big\}  \label{DE dir}
\end{equation}
where the two dimensionless functions $b_0$ and $b_2$ can be obtained from
the knowledge of the Casimir free energy for small departures
around the $\psi({\mathbf x_\parallel}) = a = {\rm constant}$ case. Indeed, 
setting:
\begin{equation}
	\psi({\mathbf x_\parallel}) \;=\; a \,+\, \eta({\mathbf
	x_\parallel}) \;,
\end{equation}
one expands $\Gamma_\beta$ in (\ref{eq:fbd}) in powers of $\eta$, up to the second order. Thus,
\begin{equation}\label{eq:expansion1}
	\Gamma_\beta(a,\eta)\;=\; \Gamma_\beta^{(0)}(a) \;+\;\Gamma_\beta^{(1)}(a,\eta) \;+\;
 \Gamma_\beta^{(2)}(a,\eta) \;+\;\ldots
\end{equation}
where the index denotes the order in $\eta$.

For the expansion above, $\Gamma_\beta^{(0)}$ is proportional to the area
of the mirrors, $L^{d-1}$. In terms of the Matsubara
frequencies: $\omega_n \equiv \frac{2\pi n}{\beta}$, $n \in {\mathbb Z}$, the explicit form of the zero
order term per unit area is as follows:
\begin{eqnarray}\label{eq:fb0}
	\frac{\Gamma_\beta^{(0)}(a)}{L^{d-1}} &=& \frac{1}{2\beta}\,
	\sum_{n=-\infty}^{+\infty} \int \frac{d^{d-1}{\mathbf p}_\parallel}{(2\pi)^{d-1}} \,
	\log\Big[ 1 - e^{ - 2 a \sqrt{\omega_n^2 + {\mathbf p}^2_\parallel}} \Big] \nonumber\\
	&=& \frac{1}{a^d} \, b_0(\frac{a}{\beta}) \;, 
\end{eqnarray}
where:
\begin{eqnarray}\label{eq:b0gen}
b_0(\xi) &=& \frac{\xi}{2}\, \sum_{n=-\infty}^{+\infty} 
\int \frac{d^{d-1}{\mathbf p}_\parallel}{(2\pi)^{d-1}} \,
\log\Big[ 1 - e^{ - 2 \sqrt{(2\pi n \,\xi)^2 + {\mathbf p}^2_\parallel}}
\Big] \nonumber\\
&=& \xi \, \frac{\pi^{(1-d)/2}}{2^{d-1} \Gamma(\frac{d-1}{2})} \,  \sum_{n=-\infty}^{+\infty} 
\int_0^\infty d\rho \rho^{d-2}\, \log\Big[ 1 - e^{ - 2 \sqrt{(2\pi n
	\,\xi)^2 + \rho^2}}\Big] 
\end{eqnarray}
is the dimensionless function which appears in the DE for the zero order
term ($\xi = a/\beta$).

Regarding $\Gamma_\beta^{(2)}$, which is necessary in order to find
$b_2$, the result can be presented in a more compact form in terms of its
Fourier space version. Defining the spatial Fourier transform of $\eta$ by:
\begin{equation}
\eta({\mathbf x_\parallel}) \,=\, \int \frac{d^{d-1}{\mathbf k}_\parallel}{(2\pi)^{d-1}} 
e^{i {\mathbf k}_\parallel \cdot {\mathbf x}_\parallel} \; {\widetilde \eta}({\mathbf k}_\parallel) 
\end{equation}
we have
\begin{equation}
	\Gamma_\beta^{(2)} \,=\, \frac{1}{2} 
	\int \frac{d^{d-1}{\mathbf k}_\parallel}{(2\pi)^{d-1}}\, 
	f^{(2)}(0, {\mathbf k}_\parallel) \,
\big|\widetilde{\eta}({\mathbf k}_\parallel)\big|^2 
\end{equation}
where 
$$
f^{(2)}(\omega_n, {\mathbf k}_\parallel) = -\frac{2}{\beta} \sum_{m=-\infty}^{+\infty} 
\int \frac{d^{d-1}{\mathbf p}_\parallel}{(2\pi)^{d-1}}
\sqrt{\omega_m^2 + {\mathbf p}^2_\parallel} \; 
\sqrt{(\omega_m + \omega_n)^2 + ({\mathbf p}_\parallel + {\mathbf
k}_\parallel)^2} \, 
$$
$$ \times  \frac{1}{1 - \exp\big(- 2 a \sqrt{\omega_m^2 + {\mathbf
p}^2_\parallel}\big)}  \; \frac{1}{\exp\big[2 a \sqrt{(\omega_m + \omega_n)^2 + ({\mathbf p}_\parallel 
+ {\mathbf k}_\parallel)^2}\big] - 1 }
$$
\begin{equation}\label{eq:fb2}
\;\equiv\;a^{-(d+2)} \; F^{(2)}( \frac{a}{\beta}; n, a|{\mathbf k}_\parallel|)  
\end{equation}
with 
\begin{eqnarray}
F^{(2)}( \xi ; n, |{\mathbf l}_\parallel|) & = & -2 \xi \, \sum_{m=-\infty}^{+\infty} 
\int \frac{d^{d-1}{\mathbf p}_\parallel}{(2\pi)^{d-1}} \Big\{
\sqrt{(2 \pi m \, \xi)^2 + {\mathbf p}^2_\parallel} 
\sqrt{(2 \pi (m +n) \, \xi)^2 + ({\mathbf p}_\parallel + {\mathbf l}_\parallel)^2} \nonumber \\
& \times & \frac{1}{1 - \exp\big[- 2 \sqrt{(2 \pi m \, \xi)^2  + {\mathbf p}^2_\parallel}\big]}
\frac{1}{\exp\big\{2 \sqrt{ [2 \pi (m + n) \, \xi]^2 + 
({\mathbf p}_\parallel + {\mathbf l}_\parallel)^2}\big\} - 1 }\Big\} \;,
\label{eq:fb3} \end{eqnarray}
which is also a dimensionless function with dimensionless arguments. We
have made explicit the fact that the result will only depend on the modulus
of ${\mathbf l}_\parallel$, as any dependence on its direction may be got
rid off by a redefinition of the integration variables.

The coefficient $b_2$ can be immediately defined in terms of $F^{(2)}$.
Indeed, 
\begin{equation}\label{eq:defb2}
b_2(\xi) \,=\, \frac{1}{2}\Big[\frac{\partial F^{(2)}(\xi ; n, |{\mathbf
l}_\parallel|)}{\partial |{\mathbf l}_\parallel|^2} \Big]_{ n\to 0, |{\mathbf
l}_\parallel| \to 0} \;.
\end{equation}

In the following subsection, we consider the low and high temperature
limits of the two coefficients $b_0$ and $b_2$; since they determine the
form of the DE in the corresponding limits. We note that the relevant scale
to compare the temperature with is the inverse of the distance between the mirrors. Thus,
in terms of the variable $\xi$, the relevant cases are: $\xi \to 0$ (zero
temperature limit) and $\xi \to \infty$ (infinite temperature limit). We
discuss them below.

\subsubsection{The zero and high temperature limits}

The zero temperature limit corresponds to $\xi \to 0$, and it can be
implemented by replacing a sum over discrete indices by an integral over
a continuous index. Defining $k_0 = 2 \pi n \xi$, we get an integral over
$k_0$; the Jacobian being $1/(2 \pi \xi)$. The results for the two
coefficients, in $d$ dimensions (we introduce the number of dimensions as
an explicit parameter), are:
\begin{equation}\label{eq:b0low}
\big[b_0(d,\xi)\big]_{\xi << 1} \,\sim\, \frac{1}{2}\, 
\int \frac{d^dp_\parallel}{(2\pi)^d} \, \log\Big[ 1 - e^{ - 2
|p_\parallel|} \Big] \equiv b_0(d) \;,
\end{equation}
and
\begin{equation}\label{eq:b2low}
\big[b_2(d,\xi)\big]_{\xi <<1}\,=\,\frac{1}{2}
\Big[\frac{\partial F^{(2)}_0(d,|l_\parallel|)}{\partial |l_\parallel|^2}
\Big]_{l_\parallel \to 0}\equiv b_2(d)\;,
\end{equation} 
where:
\begin{equation}\label{eq:b2low1}
F_0^{(2)}(d,|l_\parallel|)  =  -2 \, \int \frac{d^dp_\parallel}{(2\pi)^d}
|p_\parallel| \, |p_\parallel + l_\parallel| 
 \frac{1}{1 - e^{- 2 |p_\parallel|}}  \frac{1}{e^{2 |p_\parallel + l_\parallel|} - 1 } \;.
\end{equation}

It is possible to give a more explicit expression for $b_2(d)$, since this
coefficient may be obtained by taking derivatives inside of the integrand
of (\ref{eq:b2low1}). Also, the $b_0(d)$ coefficient can be exactly evaluated 
as a function of $d$. We present, in the following table, the ratio between
the two coefficients as a function of $d$, for $1\leq d\leq 6$:

\begin{table}
\begin{center}
\begin{tabular}{|c|c|c|}
\hline 
& $\frac{b_2(d)}{b_0(d)}$ & $\sim$ \\ 
\hline 
$d=1$ &$\frac{1}{\pi^2}(1+\frac{\pi^2}{3})$ &$0.435$ \\ 
\hline 
$d=2$ & $\frac{1+6\zeta(3)}{12\zeta(3)}$ &$ 0.569$   \\ 
\hline 
$d=3$ & $2/3$ & $0.667$ \\ 
\hline 
$d=4$ & $\frac{-\zeta (3)+10 \zeta (5)}{12 \zeta (5)}$ & $0.737$ \\ 
\hline 
$d=5$ & $\frac{10 \pi ^2-21}{10 \pi ^2}$ & $0.787$\\ 
\hline 
$d=6$ & $\frac{-2 \zeta (5)+7 \zeta (7)}{6 \zeta (7)}$ & $0.824$\\ 
\hline 
\end{tabular} 
\end{center}
\caption{Values of the ratios $\frac{b_2(d)}{b_0(d)}$ for the lowest dimensions. }
\end{table}

\begin{figure}
\centering
\includegraphics[width=8cm , angle=0]{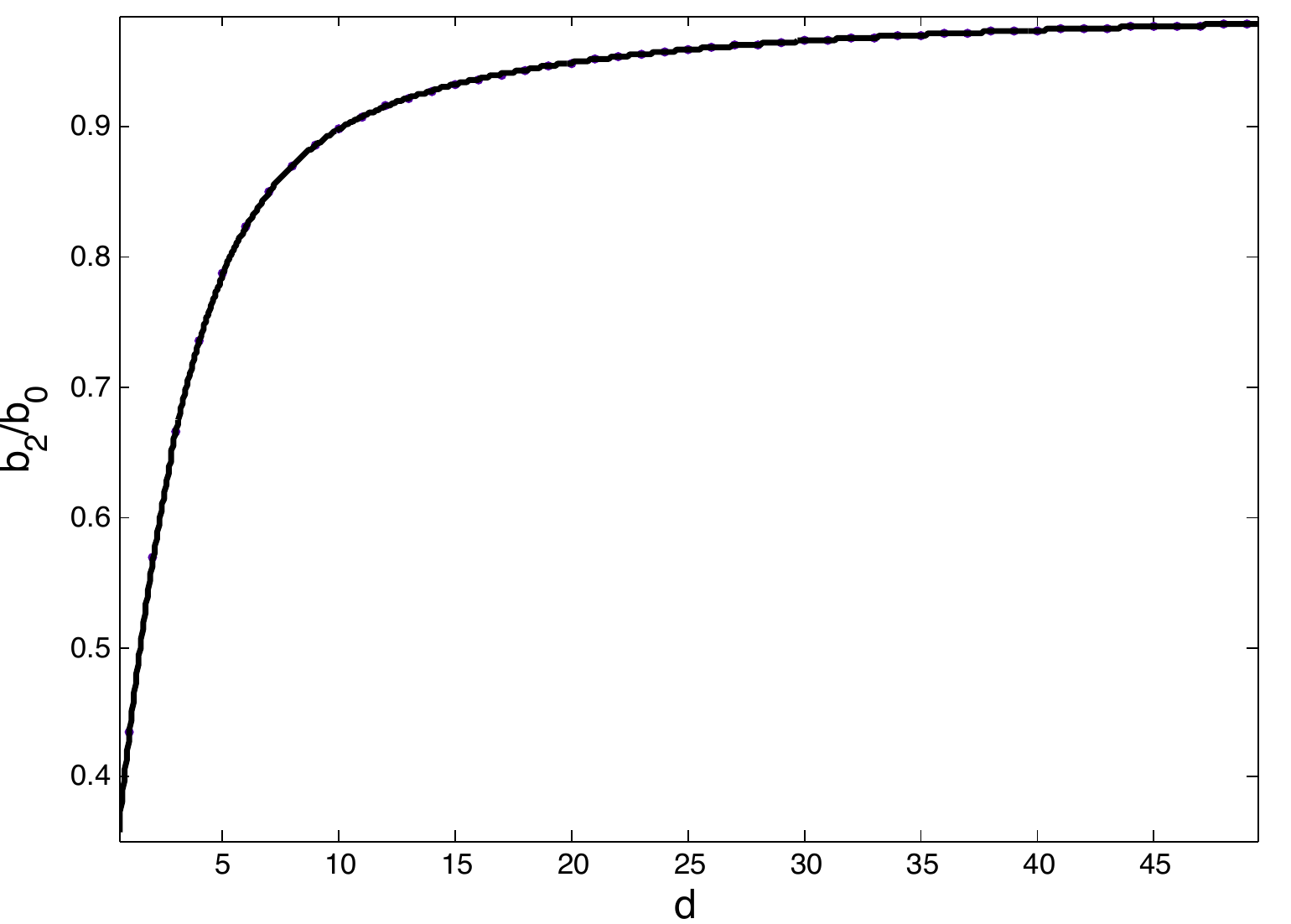}
\caption{\small{The function $b_2/b_0$ tends to $1$ for large values of $d$.}} \label{fig1}
\end{figure}

It is interesting to remark that the relative weight of the NTLO correction grows with the number of dimensions. Indeed, 
the general results  for $b_0(d)$ and $b_2(d)$ in an arbitrary number of dimensions are:
\begin{equation}
b_0(d) = -{\Gamma\left((d+1)/2\right)\,\zeta(d+1)\over (4\pi)^{(d+1)/2}}\label{b0}
\end{equation} 
and
\begin{eqnarray}
b_2(d) &=& -{1\over 12 \pi^2}{\pi^{d/2}\over 2^{d}}\Big[{(d-3)(d-1)\over d}\Gamma\left(2-{d\over 2}\right)
\zeta(2-d)\nonumber\\ 
&+& \pi^{3/2-d} (d+1)\Gamma\left({d+1\over 2}\right)\zeta(d+1)
\Big] \, \label{b2}
\end{eqnarray}
These expressions are consistent with those derived in Ref.\cite{Machado}, using a different method and in the context of the dynamical Casimir effect. Figure \ref{fig1} shows that the ratio $b_2(d)/b_0(d)$ is an increasing function
of $d$, tending to $1$ as $d\rightarrow\infty$.  


Let us now consider the very high (infinite) temperature limit.
When $\xi >> 1$, we see that only the $n=0$ term in the sum
representing $b_0$ yields a non-vanishing contribution,
\begin{equation}\label{eq:b0hight}
b_0(d,\xi) \,\sim\, \frac{\xi}{2}\, \int \frac{d^{d-1}{\mathbf p}_\parallel}{(2\pi)^{d-1}} \,
\log\big( 1 - e^{ - 2 |{\mathbf p}_\parallel|} \big) \;,
\end{equation}
or, introducing explicitly the dependence on the number of space
dimensions, $d$,
\begin{equation}\label{eq:b0high}
\big[b_0(\xi,d)]_{\xi>>1} \;\sim\; \xi \, \big[b_0(\xi, d-1)]_{\xi \to 0}
= b_0(d-1) \;, 
\end{equation}
a reflection of the well known `dimensional reduction' phenomenon at high 
temperatures, for bosonic degrees of freedom.

For the $b_2$ coefficient, a similar analysis shows that only $m=0$ has to
be kept, and: 
\begin{equation}\label{eq:b2high}
\big[b_2(\xi,d)\big]_{\xi>>1} \;\sim\; \xi \, \big[b_2(\xi, d-1)\big]_{\xi
\to 0} = b_2(d-1)\ \;. 
\end{equation}

Putting together (\ref{eq:b0high}) and (\ref{eq:b2high}), we finally get
for the DE up to the second order in the high temperature limit:
\begin{equation}\label{Fbetainf}
\big[\Gamma_\beta (\psi,d) \big]_{\psi/\beta >>1} \,\sim\, 
\frac{1}{\beta}\int d^{d-1}{\mathbf x}_\parallel  \Big\{ b_0(d-1)
\frac{1}{[\psi({\mathbf x}_\parallel)]^{d-1}} 
+ 
b_2(d-1) \frac{(\partial\psi)^2}{[\psi({\mathbf
x}_\parallel)]^{d-1}} \Big\}  \, .
\end{equation}
In particular, the free energy
reads, for $d=3$,
\begin{equation}\label{Fbetainf3}
\big[\Gamma_\beta (\psi,3) \big]_{\psi/\beta >>1} \,\sim\, 
-\frac{\zeta(3)}{16\pi\beta}\int d^2{\mathbf x}_\parallel \frac{1}{[\psi({\mathbf x}_\parallel)]^{2}} \left\{ 1+ 
 0.569\,
(\partial\psi)^2 \right\}  \, .
\end{equation}

As an example, let us now apply the results above  to the evaluation of the Casimir
interaction between a sphere and a plane at very high temperatures.  The
sphere has radius $R$, and is in front of a plane at a minimum distance $a$
($a\ll R$).  Although the surface of the sphere cannot be covered by a
single function $z = \psi(x_\parallel)$, as in previous works, we will
nevertheless consider just the region of the sphere which is closer to the
plane \cite{pfa_nos}. We shall see that this procedure still produces results which are
quantitatively adequate within the present approximation and assumptions,
even beyond the lowest order.

The function $\psi$ is
\begin{equation}
\psi(\rho) = a + R \left(1 - \sqrt{1 - \frac{\rho^2}{R^2}}\right),
\end{equation}
where we are using polar coordinates ($\rho, \phi$) for the $x_3 =0$ plane.
This function describes the hemisphere when $0\leq \rho \leq R$. The
DE will be well defined if we restrict the integrations
to the region $0\leq \rho\leq\rho_M<R$.  We will assume that $\rho_M/R=O(1)<1$.  Inserting this expression for
$\psi$ into Eq.(\ref{Fbetainf3}) and performing 
explicitly the integrations we obtain 
\begin{equation}
\big[\Gamma_\beta (\psi,3) \big]_{\psi/\beta >>1} \,\sim\, -\frac{\zeta(3)R}{8\beta a} \left(1+0.569 \frac{a}{R}\log\left(\frac{a}{R}\right)\right)\, .
\label{Fsp}
\end{equation}
Note that, as long as $a\ll R$, the force will not depend on 
$\rho_M$.  As expected on dimensional grounds, the $R/a^2$ behavior of the leading contribution in the zero temperature case changes to
$R/a\beta$ at very high temperatures.  The NTLO correction is analytic when written in terms of derivatives of the function $\psi$, but non-analytic in $\frac{a}{R}$.
This behavior has been already noted in numerical estimations of the Casimir interaction between a sphere and a plane in the infinite temperature
limit, for the electromagnetic case in Ref.\cite{Neto2012}. 


\begin{figure}
\centering
\includegraphics[width=8cm , angle=0]{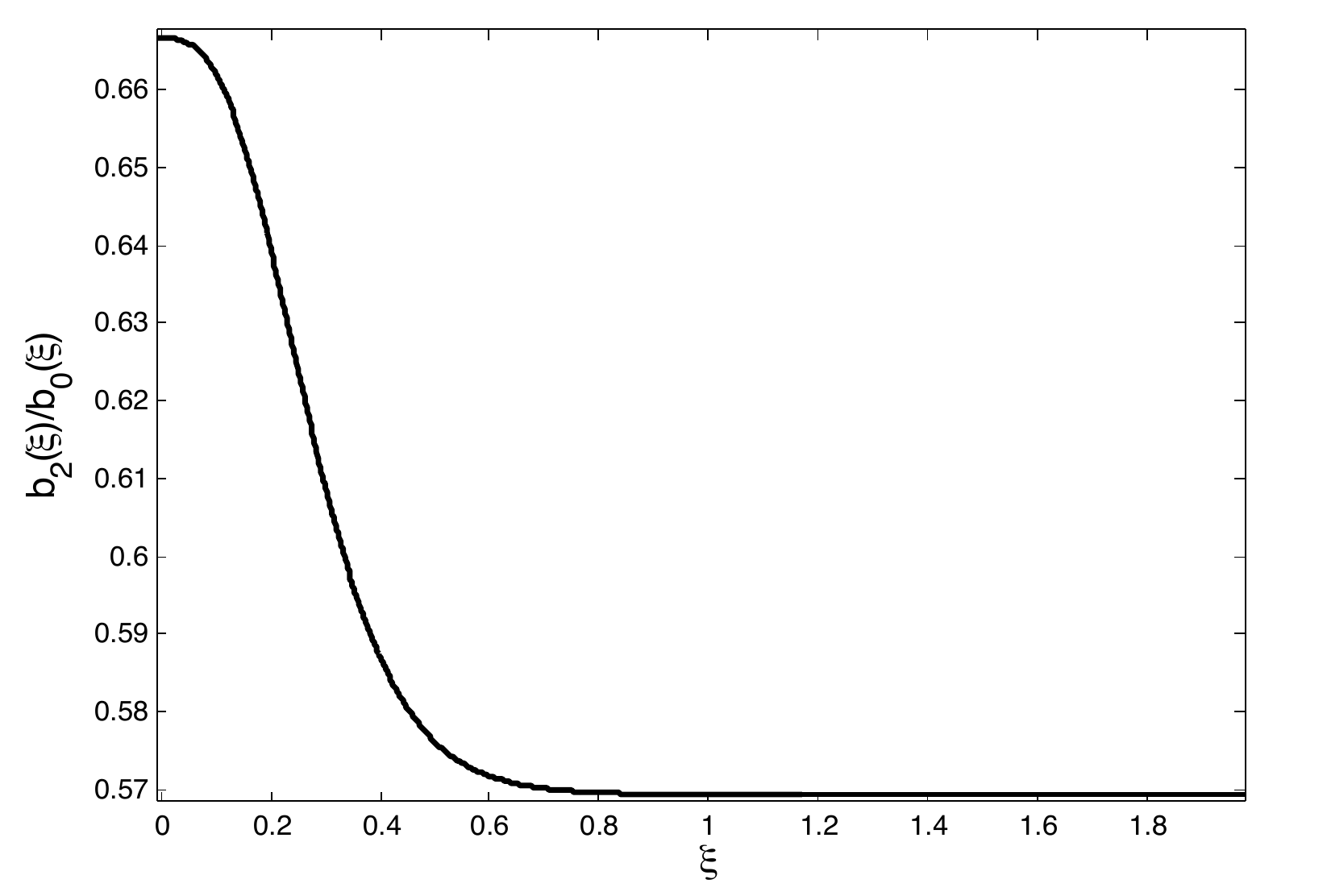}
\caption{Ratio between coefficient $b_2(\xi)$ given by Eq.(\ref{eq:defb2}) and the coefficient $b_0(\xi)$ of 
Eq.(\ref{eq:b0gen}), as a function of the dimensionless temperature $\xi$ for $d = 3$. The plot interpolates between the 
value $b_2/b_0 = 0.67$ for zero temperature, and $0.57$ at high temperatures.} \label{fig2}
\end{figure}



\begin{figure}
\centering
\includegraphics[width=8cm , angle=0]{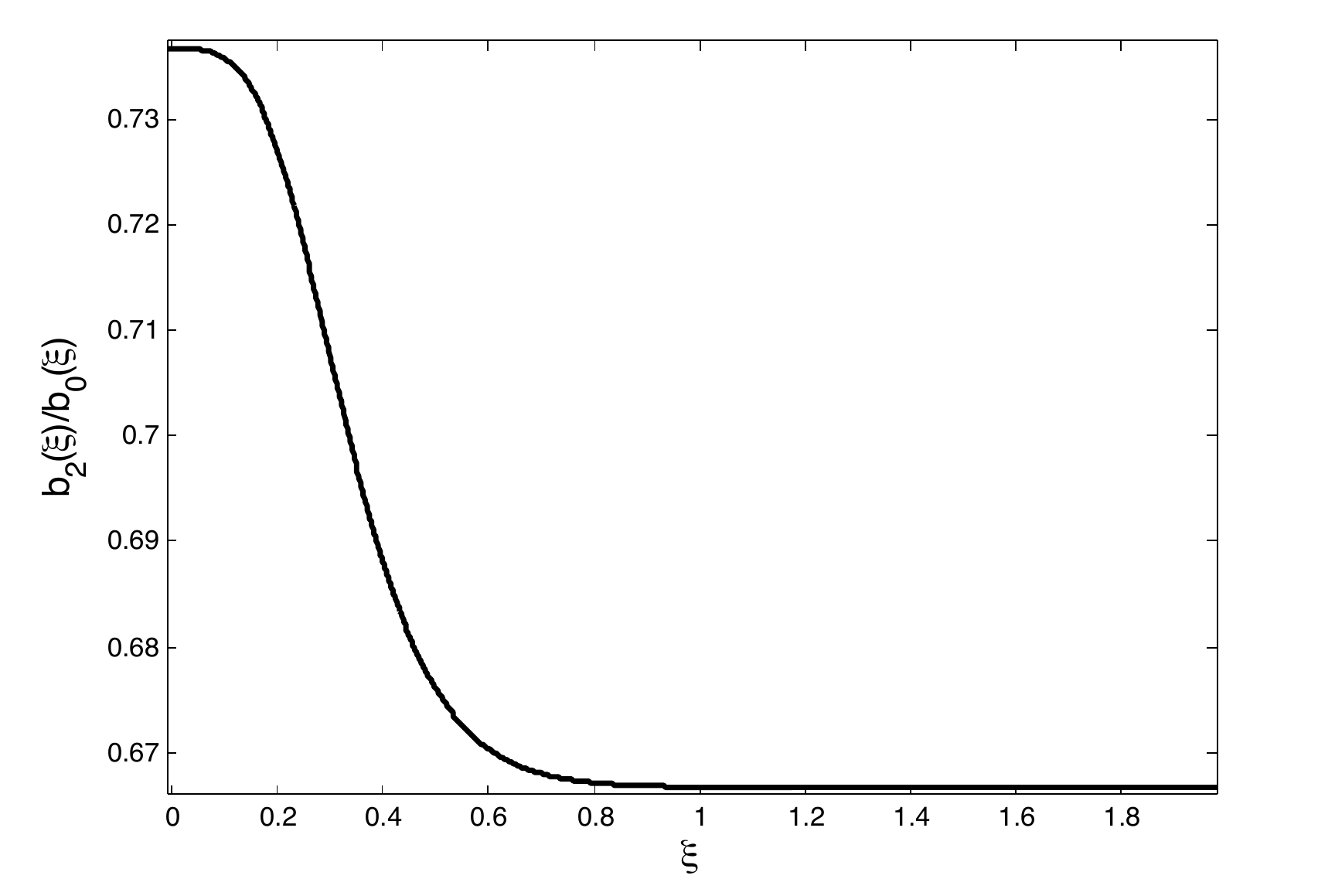}
\caption{\small{Ratio between coefficient $b_2(\xi)$ given by Eq.(\ref{eq:defb2}) and the coefficient $b_0(\xi)$ of 
Eq.(\ref{eq:b0gen}), as a function of the dimensionless temperature $\xi$ for $d = 4$. The plot interpolates between the 
value $b_2/b_0 = 0.74 $ for zero temperature, and $0.67$ at high temperatures.}} \label{fig3}
\end{figure}

It is interesting to remark that the expression for the free energy at
high temperatures in $d=3$ is quite similar to that corresponding to the
electrostatic force $F_z$ between two surfaces held at a constant potential
difference $V$ \cite{nos_annphys}
\begin{equation}
F_z\simeq - \frac{\epsilon_0V^2}{2} \int d^2x_\parallel\,  \frac{1}{\psi^2}\left[1+\frac{1}{3}(\partial\psi)^2\right]\, .
\label{FE}\end{equation}
Therefore,  when considering an arbitrary surface over a plane, the high
temperature limit of the free energy will have the same behavior  than the
electrostatic force.
For instance, from the results of \cite{nos_annphys},  for a
cylinder of radius $R$ and length $L$ at a distance $a$  of a plane, we
expect  the leading term of the free energy to be proportional to 
$\frac{L}{\beta}\sqrt{\frac{R}{a^3}}$ while its  NTLO correction must be a
coefficient times $\frac{L}{\beta}\sqrt{\frac{1}{a R}}$.

Going back to the general case, at intermediate temperatures the coefficients $b_2(\xi)$ given by Eq.(\ref{eq:defb2}) and $b_0(\xi)$ of 
Eq.(\ref{eq:b0gen}) should interpolate between their zero and high temperatures values. This is shown in Figs. \ref{fig2} and \ref{fig3},
where we plot the ratio $b_2/b_0$ as a function of the dimensionless temperature $\xi$ for $d = 3$ and $d=4$, respectively. The plot for $d=3$
interpolates between the value $b_2/b_0 = 0.67$ for zero temperature, and $0.57$ at high temperatures. On the other hand, for $d=4$,   $b_2/b_0$
interpolates between  $0.74$ for zero temperature and  $0.67$ at high temperatures. These limits are in agreement with the results  in Table 1. 
The ratio $b_2/b_0$ gives a quantitative measure of the relevance
of the NTLO correction to the PFA. Note that, both for $d=3$ and $d=4$, it converges quickly to the infinite temperature value.


\begin{figure}
\centering
\includegraphics[width=8cm , angle=0]{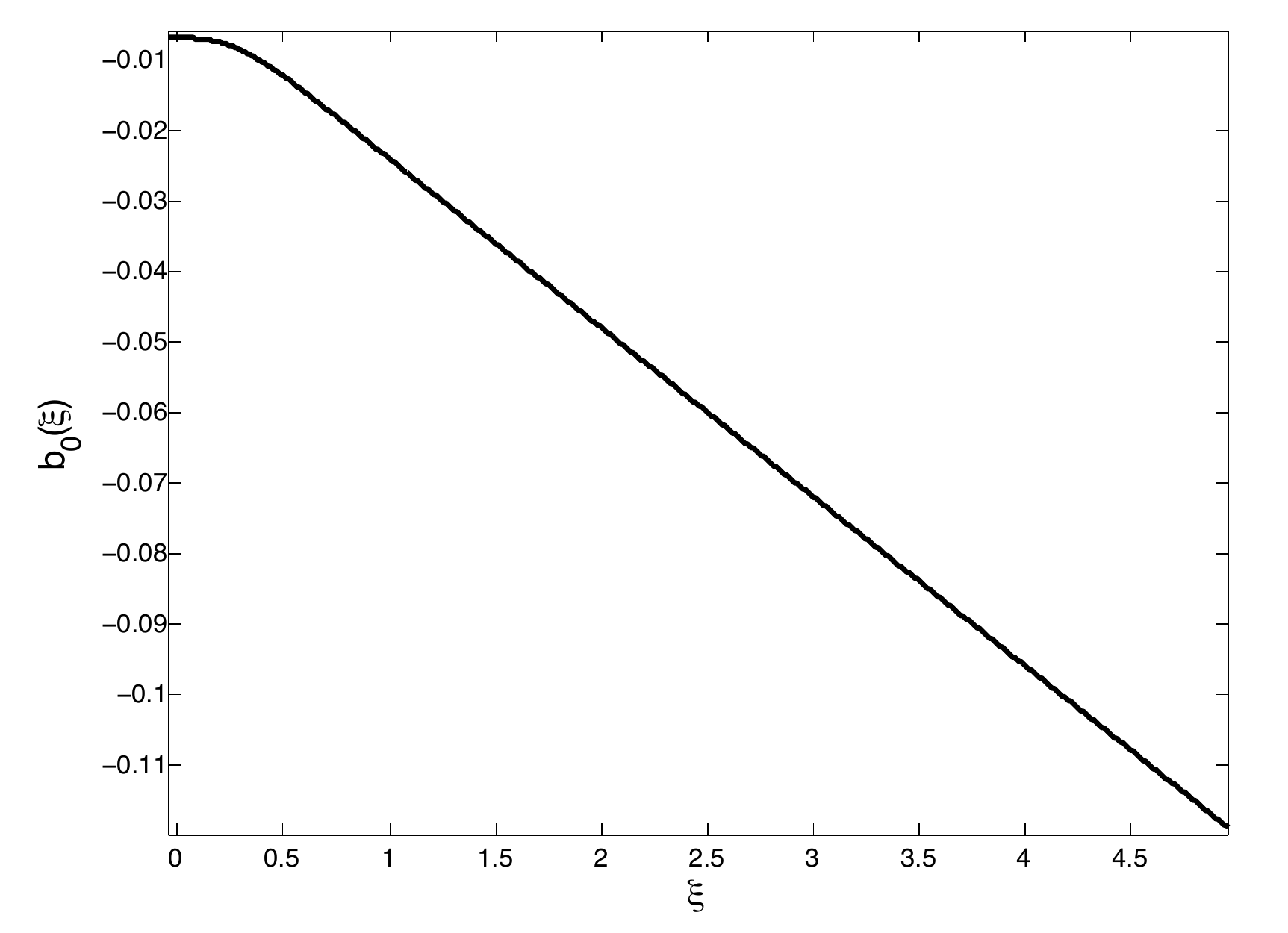}
\caption{\small{The function $b_0(\xi)$ in $3+1$ dimensions.}} \label{fig4}
\end{figure}


\begin{figure}
\centering
\includegraphics[width=8cm , angle=0]{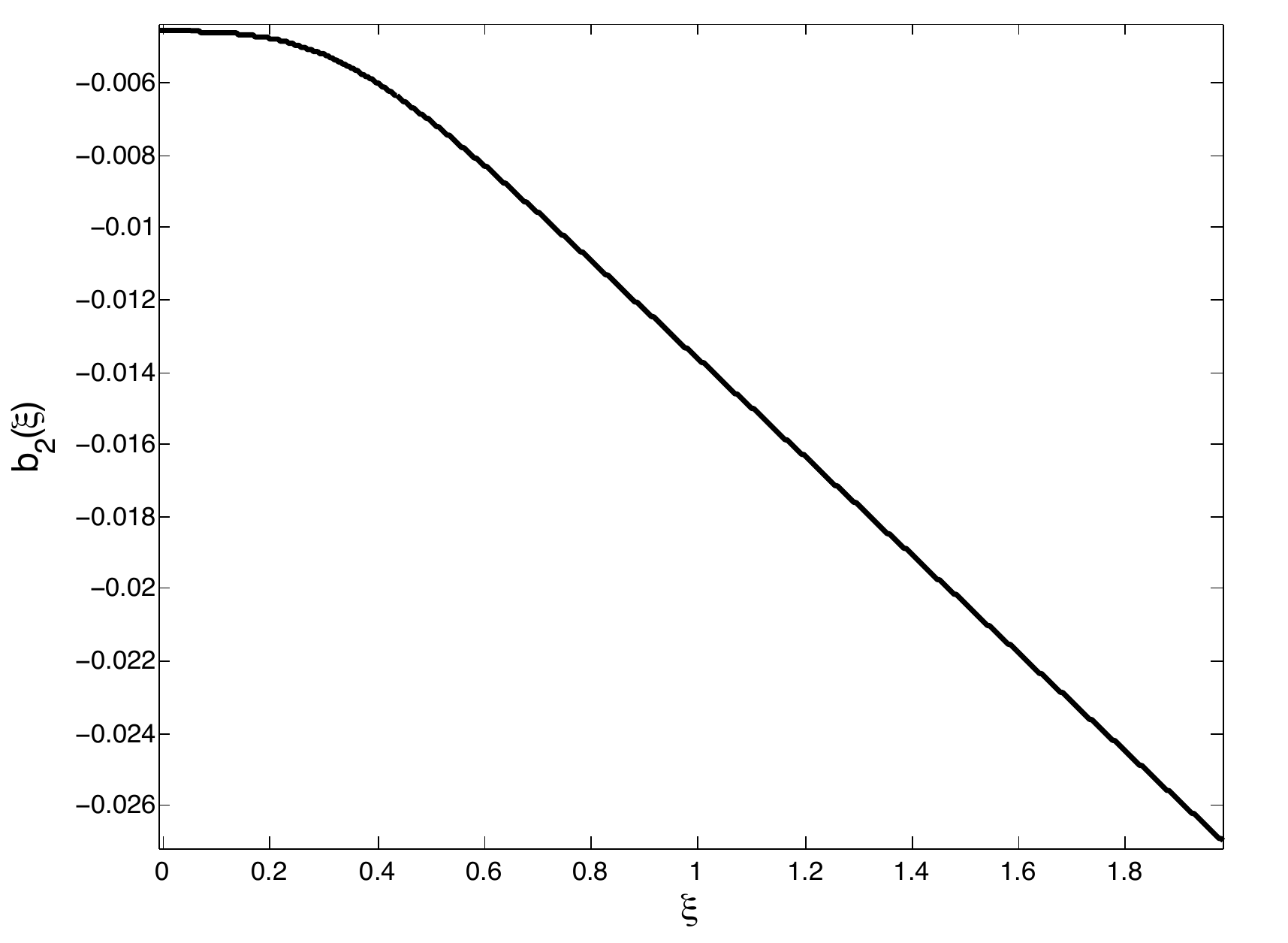}
\caption{\small{The function $b_2(\xi)$ in $3+1$ dimensions.}} \label{fig5}
\end{figure}

Finally, it is worth stressing that, as the coefficients $b_0$ and $b_2$ are functions of $\xi = T\psi$,  the evaluation of these functions is crucial in 
order to compute the Casimir free energy using the DE in any concrete example,  at a fixed temperature.  The previous plots  describe the dependence of their ratio  
with distance,  at a fixed temperature.  In Fig. \ref{fig4} we plot $b_0(\xi)$ in $3+1$ dimensions. We can see that, at low temperatures, the curve is very flat. Indeed,
it is well known that
the low temperature corrections to the free energy for parallel plates are proportional to $\xi^3$ for $\xi\ll 1$, and this behavior is well reproduced in the numerical evaluation. Moreover, the function $b_0(\xi)$ acquires very quickly the linear behavior expected at very high temperatures.  For the sake of completeness, in Fig. \ref{fig5}, we plot
the function $b_2(\xi)$, which has similar characteristics.    

These results may be useful to understand the nontrivial interplay
between geometry and temperature for open geometries, like the
sphere-plate and the cylinder-plate configurations, described in
Ref.\cite{Weber}. Indeed, it was pointed out there that local approximation 
techniques such as the PFA are generically inapplicable at low temperatures. From 
our results we see that both functions
$b_0$ and $b_2$ approach their  high temperature behavior for relatively
low values of $\xi$. Therefore  it would not be valid to insert the 
low-$\xi$ expansions of these functions  into Eq.(\ref{DE dir}), and then apply the result 
to open geometries for which the condition $\xi\ll 1$ is not satisfied.

\section{Neumann boundary conditions}\label{sec:Neumann}
Again we start from the functional representation of ${\mathcal
Z}_\beta(\psi)$ given in (\ref{eq:defzetabeta}), but now we use the 
functional $\delta$ functions which impose Neumann (rather than Dirichlet)
boundary conditions on the two mirrors. We assume the mirrors to be 
characterized by the same surfaces we used in the Dirichlet case.

The boundary conditions may be written as follows:
\begin{eqnarray}\label{eq:nbc}
&L)& \, [\partial_d \varphi(x_\parallel,x_d)]_{x_d=0} = 0 \nonumber\\
&R)& \, [\partial_n \varphi(x_\parallel,x_d)]_{x_d=\psi(x_\parallel)} = 0 \;,
\end{eqnarray}
where $\partial_n = n^\mu \partial_\mu$, with $n^\mu$ the unit normal to
the surface $x_d-\psi(x_\parallel) = 0$: 
\begin{eqnarray}
n^\mu(x_\parallel) &=& \frac{N^\mu(x_\parallel)}{|N(x_\parallel)|}
\nonumber\\ 
N^\mu(x_\parallel) &=& \delta^\mu_d  - \delta^\mu_\alpha \partial_\alpha
\psi(x_\parallel) \;,
\end{eqnarray}
and $|N(x_\parallel)|= \sqrt{g(x_\parallel)}$. 

To exponentiate the $\delta$-functionals, we again introduce two auxiliary fields, 
$\lambda_L$ and $\lambda_R$: 
\begin{eqnarray}\label{eq:deltexpn}
\delta_L(\varphi)  &=& \int {\mathcal D}\lambda_L \, 
e^{i \int d^dx_\parallel \, \lambda_L(x_\parallel) \,
[\partial_d\varphi(x_\parallel,x_d)]_{x_d=0}} 
\nonumber\\
\delta_R(\varphi)  &=& \int {\mathcal D}\lambda_R \, e^{i \int
	d^dx_\parallel \, \sqrt{g (x_\parallel)} \, \lambda_R(x_\parallel) \, 
       [\partial_n \varphi(x_\parallel,x_d)]_{x_d=\psi(x_\parallel)} }
\nonumber\\
&=& \int {\mathcal D}\lambda_R \, e^{i \int
	d^dx_\parallel \, \lambda_R(x_\parallel) \, 
       [\partial_N \varphi(x_\parallel,x_d)]_{x_d=\psi(x_\parallel)} } \; ,
\end{eqnarray}
where we introduced the notation $\partial_N=N^\mu\partial_\mu$.
Thus, using those exponential representations, we derive:
\begin{equation}\label{eq:zetanbeta1}
	{\mathcal Z}_\beta(\psi)\;=\; \int {\mathcal D}\varphi \, {\mathcal
	D}\lambda_L {\mathcal D}\lambda_R \; e^{-{\mathcal S}_0(\varphi)
	\,+\, i \int d^{d+1}x J_N(x) \varphi(x)} \;,
\end{equation}
where, by analogy with the Dirichlet case, we introduce the `current' 
$J_N(x)$:
\begin{equation}
	J_N(x) \;=\; \lambda_L(x_\parallel) \, \partial_d\delta(x_d) \,+\, 
	\lambda_R(x_\parallel) \, \partial_N \delta(x_d - \psi(x_\parallel))  \;.
\end{equation}
Note that there is no need to get rid now of any metric-dependent factor, as we did
for the Dirichlet case. 

The integral over $\varphi$ becomes then:
\begin{equation}\label{eq:zetanbeta2}
	{\mathcal Z}_\beta(\psi)\;=\; {\mathcal Z}_\beta^{(0)} \,
	\int {\mathcal D}\lambda_L {\mathcal D}\lambda_R \;
	e^{- \frac{1}{2} \int_{x_\parallel,x'_\parallel}
		\lambda_A(x_\parallel) {\mathbb U}_{AB}(x_\parallel,x'_\parallel)
	\lambda_B(x'_\parallel) }, 
\end{equation}
where:
\begin{eqnarray}
{\mathbb U}_{LL}(x_\parallel,x'_\parallel) &=& - \int
\frac{d^dk_\parallel}{(2\pi)^d} \; e^{i k_\parallel \cdot
(x_\parallel-x'_\parallel)} \;  \frac{|k_\parallel|}{2}\nonumber\\
{\mathbb U}_{LR}(x_\parallel,x'_\parallel) &=& - \int 
\frac{d^dk_\parallel}{(2\pi)^d} \; e^{i k_\parallel \cdot
(x_\parallel-x'_\parallel)} \; e^{-|k_\parallel| \psi(x'_\parallel)} 
\frac{|k_\parallel| + i k_\parallel \cdot \partial\psi(x'_\parallel)}{2} 
\nonumber\\
{\mathbb U}_{RL}(x_\parallel,x'_\parallel) &=& - \int 
\frac{d^dk_\parallel}{(2\pi)^d} \; e^{i k_\parallel \cdot
(x_\parallel-x'_\parallel)} \; e^{-|k_\parallel| \psi(x_\parallel)} 
\frac{|k_\parallel| - i k_\parallel \cdot \partial\psi(x_\parallel)}{2} 
\nonumber\\
{\mathbb U}_{RR}(x_\parallel,x'_\parallel) &=&\int \frac{d^dk_\parallel}{(2\pi)^d} \; 
e^{i k_\parallel \cdot (x_\parallel-x'_\parallel)} \;
e^{-|k_\parallel|[\psi(x_\parallel) -\psi(x'_\parallel)]} \nonumber\\
&\times& \frac{1}{2} \, \Big\{ -|k_\parallel| -  i k_\parallel \cdot 
[\partial\psi(x_\parallel)+\partial\psi(x'_\parallel)] 
+ \frac{1}{|k_\parallel|} (k_\parallel \cdot \partial\psi(x_\parallel)  k_\parallel \cdot
\partial\psi(x'_\parallel)) 
\Big\}
\end{eqnarray}
The free energy $\Gamma_\beta(\psi)$ is then
\begin{equation}\label{eq:fbn}
	\Gamma_\beta(\psi) \;=\; \frac{1}{2 \beta} {\rm Tr} \log {\mathbb U} \;,
\end{equation}
which, as in the Dirichlet case, does contain `self-energy' contributions,
to be discarded here by the same reason as there. Again, $\Psi$ is assumed
to be time independent.

\subsection{Derivative expansion}\label{ssec:nderivative}
Assuming that one could proceed as in the Dirichlet case, keeping up to two
derivatives, the derivative expanded Casimir free energy could be written
as follows:
\begin{equation}
	\Gamma_\beta (\psi) \;=\; \int d^{d-1}{\mathbf x}_\parallel \, 
\Big\{ c_0(\frac{\psi}{\beta}) \frac{1}{[\psi({\mathbf x}_\parallel)]^d} 
\,+\, 
c_2(\frac{\psi}{\beta}) \, \frac{(\partial\psi)^2}{[\psi({\mathbf x}_\parallel)]^d} 
\Big\}  
\end{equation}
with two new dimensionless functions $c_0$ and $c_2$. Those coefficients
may be determined from the knowledge of the Neumann Casimir free energy for small departures
around the $\psi({\mathbf x}_\parallel) = a = {\rm constant}$ case, up to
the second order in the departure. Again, we focus on the cases of purely
quantum or purely thermal effects, except for the more realistic case of
$d=3$. 
As we will  show in what follows, the NTLO term is quadratic,
except when $d=2$ at zero temperature, or when $d=3$ and there is a non
zero (finite or infinite) temperature. 

We present the calculation of the terms
contributing to that expansion in Appendix A.

It is quite straightforward to see that the zero order term coincides with
the one for the Dirichlet case; namely: $c_0 = b_0$.

$\Gamma_\beta^{(2)}$ has the following form:
\begin{equation}
	\Gamma_\beta^{(2)} \,=\, 
\frac{1}{2} \int \frac{d^{d-1}{\mathbf k}_\parallel}{(2\pi)^{d-1}}\, 
	g^{(2)}(0, {\mathbf k}_\parallel) \,
\big|\widetilde{\eta}({\mathbf k}_\parallel)\big|^2 
\end{equation}
where 
$$
g^{(2)}(\omega_n, {\mathbf k}_\parallel) = -\frac{2}{\beta} \sum_{m=-\infty}^{+\infty} 
\int \frac{d^{d-1}{\mathbf p}_\parallel}{(2\pi)^{d-1}}
\frac{ [\omega_m (\omega_m + \omega_n) + {\mathbf p}_\parallel \cdot
({\mathbf p}_\parallel + {\mathbf k}_\parallel)]^2}{\sqrt{\omega_m^2 + {\mathbf p}^2_\parallel}  
\sqrt{(\omega_m + \omega_n)^2 + ({\mathbf p}_\parallel + {\mathbf
k}_\parallel)^2}}  
$$
$$
\times \frac{1}{1 - \exp\big(- 2 a \sqrt{\omega_m^2 + {\mathbf
p}^2_\parallel}\big)}  \; \frac{1}{\exp\big[2 a \sqrt{(\omega_m + \omega_n)^2 + ({\mathbf p}_\parallel 
+ {\mathbf k}_\parallel)^2}\big] - 1 } 
$$
\begin{equation}\label{eq:gb2}
= a^{-(d+2)} \; G^{(2)}( \frac{a}{\beta}; n, a|{\mathbf k}_\parallel|)  
\end{equation}
with 
\begin{eqnarray}
G^{(2)}( \xi ; n, |{\mathbf l}_\parallel|) & = & -2 \, \xi \, \sum_{m=-\infty}^{+\infty} 
\int \frac{d^{d-1}{\mathbf p}_\parallel}{(2\pi)^{d-1}}
 \frac{ [(2 \pi \xi)^2 m (m + n) + {\mathbf p}_\parallel \cdot
({\mathbf p}_\parallel + {\mathbf l}_\parallel)]^2}{\sqrt{ (2 \pi \xi)^2 m^2 + {\mathbf p}^2_\parallel}  
\sqrt{ (2\pi\xi)^2 (m +n)^2 + ({\mathbf p}_\parallel + {\mathbf l}_\parallel)^2}}  \nonumber \\
& \times & \frac{1}{1 - \exp\big(- 2 \sqrt{(2 \pi \xi)^2 m^2 + {\mathbf
p}^2_\parallel}\big)} 
\frac{1}{\exp\big[2 \sqrt{(2\pi\xi)^2 (m +n)^2  + ({\mathbf p}_\parallel 
+ {\mathbf l}_\parallel)^2}\big] - 1 } \;. \label{eq:gb3}
\end{eqnarray}
\begin{equation}\label{eq:defbn2}
c_2(\xi) \,=\, \frac{1}{2} \Big[\frac{\partial G^{(2)}(\xi ; n, |{\mathbf
l}_\parallel|)}{\partial |{\mathbf l}_\parallel|^2} \Big]_{n\to 0, |{\mathbf
l}_\parallel| \to 0} \;. 
\end{equation}

\subsubsection{The zero and high temperature limits}

As before, the zero temperature limit can be
implemented by replacing a sum over discrete indices by an integral over
a continuous index. The result for the 
coefficient $c_2$ in $d$ dimensions is:
\begin{equation}\label{eq:c2low}
\big[c_2(d,\xi)\big]_{\xi <<1}\,=\,
\Big[\frac{\partial G_0^{(2)}(|l_\parallel|)}{\partial |l_\parallel|^2}
\Big]]_{l_\parallel \to 0}\equiv c_2(d)\;,
\end{equation} 
where:
\begin{equation}\label{eq:c2low1}
G_0^{(2)}(d,|l_\parallel|)  = 
 - 2\, \int \frac{d^dp_\parallel}{(2\pi)^d}
	\frac{\big[p_\parallel \cdot (p_\parallel + l_\parallel)\big]^2}{|p_\parallel| \, 
	|p_\parallel + l_\parallel|} \,
\frac{1}{1 - e^{- 2 |p_\parallel|}}  \frac{1}{e^{2 |p_\parallel + l_\parallel|} - 1 } \;.
\end{equation}

For $d=1$, the coefficient $c_2$ coincides with its Dirichlet counterpart $b_2$. In higher dimensions, the structure of the form factor is different.
We present, in Table 2, the ratio between $c_2(d)$ and $c_0(d)\equiv b_0(d)$
as a function of $d$, for $d\neq 2$:

\begin{table}
\begin{center}
\begin{tabular}{|c|c|c|}
\hline 
& $\frac{c_2(d)}{c_0(d)}$ & $\sim$ \\ 
\hline 
$d=1$ &$\frac{1}{3}- \frac{\zeta(0)}{3\zeta(2)}$ &$0.435$ \\ 
\hline 
$d=3$ & $\frac{2}{3}-\frac{4\zeta(2)}{3 \zeta(4)}$ & $-1.360$ \\ 
\hline 
$d=4$ & $\frac{5}{6}-\frac{19\zeta(3)}{12\zeta(5)}$ & $-1.002$ \\ 
\hline 
$d=5$ & $1-\frac{9\zeta(4)}{5\zeta(6)}$ & $-0.915$\\ 
\hline 
$d=6$ & $\frac{7}{6}-\frac{2\zeta(5)}{\zeta(7)}$ & $-0.890$\\ 
\hline 
$d=7$ & $\frac{4}{3}-\frac{46 \zeta(6)}{\zeta(8)}$ & $-0.886$\\ 
\hline 

\end{tabular} 
\end{center}
\caption{Values of the ratios $\frac{c_2(d)}{c_0(d)}$ for the lowest dimensions. Note that we have excluded the case $d=2$.}
\end{table}

We see that  the ratio  $c_2(d)/c_0(d)$ is non-monotonous and always negative for $d\neq 1$. We have also checked 
that $c_2(d)/c_0(d) \to -1$ for large values of $d$.

In the particular case $d=2$   is not possible to compute the coefficient
by introducing the derivative with respect to $\vert l_\parallel\vert^2$
inside the integral in Eq.(\ref{eq:c2low1}), because of infrared
divergences.  This is a signal of a branch cut at zero momentum, as we will
show in Section~\ref{ssec:nonanalit}.

The high temperature limit can be obtained,  as for the Dirichlet case, taking the limit  $\xi >> 1$. ``Dimensional reduction" takes place, and the free energy becomes
\begin{equation}\label{FbetainfNeu}
\big[\Gamma_\beta (\psi,d) \big]_{\psi/\beta >>1} \,\sim\, 
\frac{1}{\beta}\int d^{d-1}{\mathbf x}_\parallel  \Big\{ b_0(d-1)
\frac{1}{[\psi({\mathbf x}_\parallel)]^{d-1}} 
+ 
c_2(d-1) \frac{(\partial\psi)^2}{[\psi({\mathbf
x}_\parallel)]^{d-1}} \Big\}  \, .
\end{equation}


\begin{figure}
\centering
\includegraphics[width=8.2cm , angle=0]{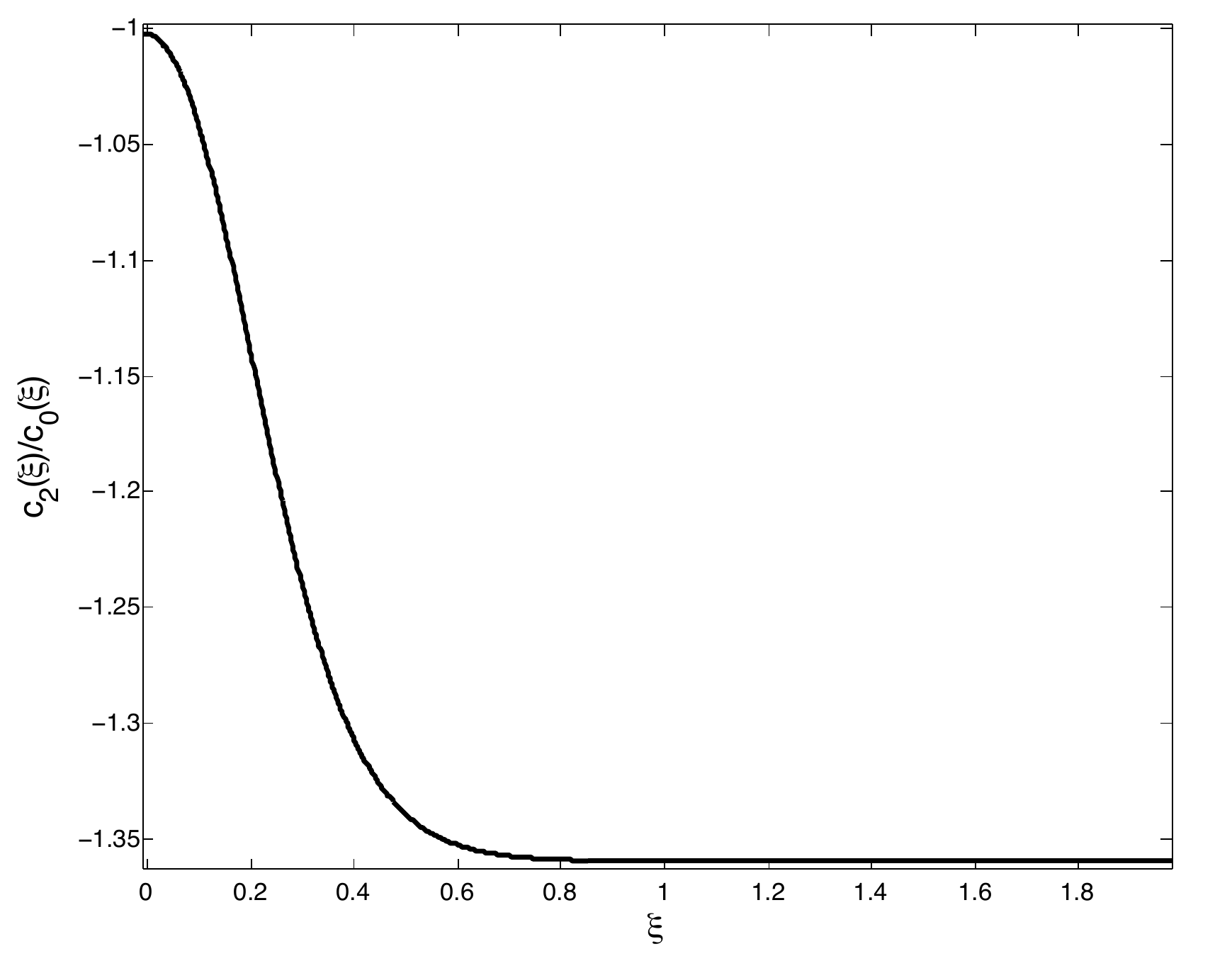}
\caption{Ratio between coefficient $c_2(\xi)$ and the coefficient $c_0(\xi)$, 
as a function of the dimensionless temperature $\xi$, for the Neumann boundary condition in $d = 4$ dimensions. The plot 
interpolates between the value $c_2/c_0 = -1.00$ and $-1.35$ at zero and high temperatures respectively.} \label{fig6}
\end{figure}

Regarding intermediate temperatures, Fig. \ref{fig6} shows the ratio between coefficients $c_2(\xi)$ and $c_0(\xi)$, 
as a function of the dimensionless temperature $\xi$, for the Neumann boundary condition in $d = 4$ dimensions. The plot 
interpolates between the value $c_2/c_0 = -1.00$ and -1.36 at zero and high temperatures respectively, in agreement with the results shown in Table 2.

\subsection{Non analytic terms:  $d=2$ with $T=0$,  and  $d=3$ with $T>0$}\label{ssec:nonanalit}

Let us consider the particular case of $d=2$ at zero temperature. As shown
in Appendix A,  for small departures of the plane-plane geometry
$\psi({\mathbf x}_\parallel)=a+\eta({\mathbf x}_\parallel)$, the correction
to the Casimir energy  reads, up to second order in $\eta$
\begin{equation}
	\Gamma_\infty^{(2)} \;=\; \frac{1}{2} \, \int
	\frac{d^2k_\parallel}{(2\pi)^2} \,
	\big[g^{(2)}(k_\parallel)\big]_{k_0 \to 0} \,
	\big| \tilde \eta({\mathbf k}_\parallel) \big|^2 \;,
\end{equation}
with:
\begin{equation}
g^{(2)}(k_\parallel)  = - 2\, \int \frac{d^2p_\parallel}{(2\pi)^2}
	\frac{\big[p_\parallel \cdot (p_\parallel + k_\parallel)\big]^2}{|p_\parallel| \, 
	|p_\parallel + k_\parallel|} \, 
\frac{1}{1 - e^{- 2 a |p_\parallel|}}  \frac{1}{e^{2a |p_\parallel + k_\parallel|} - 1 } \;.
\label{g22}
\end{equation}
Naively, one would expect the form factor $g^{(2)}(k_\parallel)$ to admit an expansion in powers of $k_\parallel^2$, which is the necessary condition
in Fourier space to produce a DE in configuration space.  However, this is not the case for $d=2$, as suggested by the fact that the formal
expression  
\begin{equation}
\frac{\partial g^{(2)}}{\partial k_\parallel^2}\vert_{k_\parallel\to 0}=- 2\, \int \frac{d^2p_\parallel}{(2\pi)^2}\frac{1}{|p_\parallel| \, (1 - e^{- 2 a |p_\parallel|})}
\frac{\partial}{\partial k_\parallel^2}\left[
\frac{\big[p_\parallel \cdot (p_\parallel + k_\parallel)\big]^2}{|p_\parallel| \, 
	|p_\parallel + k_\parallel|} \, 
 \frac{1}{e^{2 a |p_\parallel + k_\parallel|} - 1 }\right]_{k_\parallel\to 0} 
\end{equation} 
has an infrared logarithmic divergence at $p_\parallel=0$. 

The behavior of $g^{(2)}(k_\parallel)$ for small values of $k_\parallel$ can be determined by studying the integral that defines it in Eq.(\ref{g22}) in the region
$a\vert p_\parallel\vert\ll 1$. In this region, and assuming also that $a\vert k_\parallel\vert\ll 1$ one can make the approximation
\begin{equation}
\frac{1}{e^{\pm 2 x}-1}\approx \pm \frac{1}{2x}
\end{equation}
and compute the integrals analytically.  In this way, it is possible to show that
\begin{equation}
g^{(2)}(k_\parallel)\approx g^{(2)}(0)-\frac{k_\parallel^2}{16\pi a^2}\log(k_\parallel^2a^2)+
O(k_\parallel^2/a^2)\, .
\end{equation}
This behavior of $g^{(2)}$, that we confirmed with a numerical evaluation of Eq.(\ref{g22}),  shows that a {\it local} DE breaks down for Neumann boundary conditions at $d=2$. However, one can still perform an expansion
for smooth surfaces, including nonlocal contributions in the Casimir energy. For instance, in the present case, the NTLO correction to the PFA will be nonlocal and  proportional to
\begin{equation}
\int d^2x_\parallel\eta(x_\parallel)\nabla_\parallel^2\log(-a^2\nabla_\parallel^2)\eta(x_\parallel)\, .
\end{equation}
As we will describe more generally in the next Section,  the breakdown of
the local expansions is related to the existence of massless modes in the
theory. These modes are generally allowed by Neumann but not for Dirichlet
boundary conditions, that impose a mass gap of order $1/a$.

The logarithmic behavior of the form factor in $d=2$  induces a similar
non-analiticity for $d=3$ at finite temperature. Indeed, the $m=n=0$ term
in the finite temperature form factor given in Eq.(\ref{eq:gb3}) is
formally identical to the Neumann form factor $g^{(2)}$ in $d=2$.
Therefore, in an expansion for small values of $\vert k_\parallel\vert$, in addition to  a term proportional to $k_\parallel^2$, 
there is a contribution proportional to $(Ta) k_\parallel^2
\log(k_\parallel^2a^2)$ at any non-vanishing temperature, which is not
cancelled by the rest of the sum over Matsubara frequencies.  

\section{Higher order terms in the DE}
\label{sec:hoDE}


In this Section we discuss some general aspects of the derivative
expansion, related with the calculation of higher orders and the eventual
breakdown of the expansion. 

In this and previous works we considered just the NTLO correction to the
PFA, which contains up to two derivatives of the
function $\psi$ in the free energy.  We expect the next to NTLO (NNTLO)  order
corrections to contain terms of the form
\begin{equation}
\vert\partial\psi\vert^4, \, \psi\vert\partial\psi\vert^2\partial^2\psi, \,\psi^2
\partial^2\psi\partial^2\psi,\,  \psi^2 \partial_\alpha \partial_\beta \psi \partial_\alpha \partial_\beta \psi, \, \psi^3 \partial^2\partial^2 \psi\, ,
\end{equation}
and terms containing more derivatives  for higher orders.

The main question to be answered is whether the free energy can be expanded
or not in local terms up to any desired order. For this to hold true, a
necessary condition is that the expansion must hold true for a particular case: 
when $\psi=a+\eta$, with $\eta\ll 1$, and one keeps just the quadratic term
in $\eta$. 

For simplicity, we deal with the $T=0$ case in $d=3$ dimensions, an example that will
illuminate several aspects of the problem.
From the previous sections, we see that the quadratic contribution to
the Casimir energy has the form:
\begin{equation}
	\Gamma_\infty^{(2)}\,=\, \frac{1}{2 a^5}\int \frac{d^2{\mathbf k}_\parallel}{(2\pi)^2}\, 
	h^{(2)}(0, a{\mathbf k}_\parallel) \,
\big|\widetilde{\eta}({\mathbf k}_\parallel)\big|^2 \, ,\label{ecas3d}
\end{equation}
where the form factor $h^{(2)}$ depends on the boundary conditions of the
quantum field. In most of the paper we considered the case of static
surfaces, but here it will be useful to  discuss the more general case in which the
right mirror can be in motion. In this situation,
on
general grounds, we expect the form factor to be a function of $a(k_0^2+
{\mathbf k}_\parallel^2)^{1/2}\equiv a\vert k_\parallel\vert $, and of course the
explicit calculations  confirm this fact. Within these approximations, the
Casimir energy will not admit an expansion in derivatives if the
form factor includes, for instance,  odd powers or logs of its argument.
Note that at higher orders, and relaxing the condition of a quadratic
approximation in $\eta$ new non-analytic terms may arise, which can be of
the same order in the DE as the ones that come from the term quadratic in
$\eta$.

In 3+1 dimensions, this question can be answered from the explicit
expressions of the form factors \cite{Emig} presented in Appendix B.
For Dirichlet
boundary conditions, the expansion of the form factor contains, in addition
to even powers of the argument, a term proportional  to $a^5\vert
k_\parallel \vert^5$ (see Eq.(\ref{seriesgtm})).
The non-analytic term becomes a nonlocal contribution in
configuration space, that cannot be expanded in derivatives of $\eta$. Note
that this contribution does not depend on the distance between mirrors.
There is a simple interpretation of this term: when considering a flat
moving  boundary, photon creation produce an imaginary part in the  vacuum
persistence amplitude when rotated from Euclidean to Minkowski spacetime.
Therefore, the expression for $\Gamma_\infty^{(2)}$ cannot be analytic in
$k_\parallel^2$.  For a
single nonrelativistic mirror, this will lead to a dissipative force
proportional to the fifth time-derivative of the position. Indeed, the
proper analytic continuation of the  Euclidean term proportional to $\vert
k_0\vert ^5$ to real time, produces this dissipative force. For the Dirichlet
case, there are no non-analyticities dependent on the distance between
mirrors. Physically, this is due to the fact that there is a frequency
threshold to produce real photons {\it between} mirrors, which is of order
$1/a$. The conclusion is that, for Dirichlet boundary conditions, the
$a-$dependent part  of the DE is well defined up to any
order.  On the other hand, the $a$-independent part contains a non-analytic
contribution related to the possibility of creating photons of arbitrary
low energies from the vacuum. 

The situation changes for the case of Neumann boundary condition.
As shown in Eq.(\ref{seriesgte}),  in
addition to the non-analytic term proportional to  $\vert k_\parallel\vert^5a^5$,
which as before produce a contribution to $\Gamma_\infty^{(2)}$ that is independent of
$a$, there is a term proportional to $\vert k_\parallel\vert^3a^3$.  The physical
reason of the existence of this term is again clear using the connection
with the DCE. Indeed, this term comes from the possibility of creating TM
photons between mirrors, moving parallel to the mirrors, for which there is
no energy  threshold. If the DE is used to compute the
force on a moving mirror, this $a$-dependent dissipative contribution to
the force will be missed, i.e. it is only possible to get the dispersive
part of the force. On the other hand, for a static and non-flat mirror,
this term produces a nonlocal component of the force, which will be
smaller than the NTLO, but the dominant part of the NNTLO.  

It is interesting to remark that both the non-analytic contributions
proportional to  $\vert k_\parallel \vert ^5$ for Dirichlet and Neumann boundary
conditions in $3+1$ dimensions, and the $\vert k_\parallel\vert ^3$ term for the
Neumann case, could be derived from the form factors described in the
previous section by analyzing the infrared behavior of the integrals in
$p_\parallel$, as we did in Sec. \ref{ssec:nonanalit}. Moreover,  it is
clear that all of them have the same physical origin: the existence of
massless degrees of freedom. For Neumann boundary conditions at $T=0$,  the
non-analyticities show up in the NTLO for $d=2$, and in the NNTLO for
$d=3$. For Dirichlet boundary conditions the non-analytic term is
independent of the distance between mirrors, and therefore does not
contribute to the Casimir force. 

 The situation is analogous to that of effective field theories that
involve massless particles. In that case, in addition to local terms in the
effective action, there are nonlocal (or non-analytic contributions) which
can be interpreted as arising from the fact that there is no threshold for
creating such particles \cite{Donoghue}. A prototypical example is quantum
field theory under the influence of external (classical) backgrounds. For massive quantum fields
in curved spaces  \cite{BooksQFTCS}, the effective action and the energy momentum tensor of
the quantum fields can be approximated by a DE  (usually
known as the Schwinger DeWitt expansion in that context). Each subsequent
term in the expansion contains additional derivatives of the metric and
inverse powers of the mass of the quantum field. The expansion is valid as
long as the typical scale of variation of the classical background is much larger
than the inverse mass. However, for massless quantum fields, it is
necessary to consider nonlocal contributions. In our case, the role   of
the background is played by the curved surface, and there are both massive
and massless excitations: the massive ones are the Dirichlet modes (TE photons) inside the
"cavity". The massless ones are the Neumann modes (TM photons) with momentum in the
direction parallel to the plates, and TE and TM photons outside the cavity.  

The physical picture suggests possible ways out to improve the PFA even
beyond the NTLO correction. This would involve a separate treatment of
massless and massive excitations. We hope to address this issue in a future
work.
 
\section{Conclusions}\label{sec:conclusions}

We have obtained explicit expressions for the NTLO term in a DE for the
Casimir free energy for a real scalar field in $d$ spatial dimensions.  The field
satisfies either Dirichlet or Neumann boundary conditions on two static
mirrors, one of them flat and located at the $x_d=0$ plane, while the other
is described by the equation: $x_d=\psi(x_1,x_2,\ldots,x_{d-1})$.  We have shown
that, for Dirichlet boundary conditions, the NTLO term in the Casimir
energy is always of quadratic order in derivatives, regardless of the number of
dimensions. Therefore it is local, and determined by a single coefficient.
We evaluated the ratio between that coefficient and the one for the PFA
term, for different values of $d$ at zero and high temperatures.   

We have also shown that the same holds true, if $d \neq 2$, for a  field
which satisfies Neumann conditions. When $d=2$, the NTLO term becomes
nonlocal in coordinate space, which is a clear manifestation of the
existence of gapless excitations allowed by the Neumann conditions \cite{footnote}. It may be seen that among all
the possible combination of linear boundary conditions on the mirrors, just this
case, Neumann conditions on both mirrors, can produce these modes. 
 
When including thermal effects, we have shown that, for Dirichlet mirrors,
the NTLO term in the free energy is also well defined (local) for any
temperature $T$. Besides, it interpolates between the proper limits:
namely, when  $T \to 0$ it tends to the one we had calculated for the 
Casimir energy, while for $T \to \infty$ it
corresponds to the one for a $d=2$ theory, realizing the expected dimensional
reduction at high temperatures.  
On the contrary, for Neumann mirrors in $d=3$, we found a nonlocal NTLO term for any 
$T>0$, which vanishes linearly when $T \to 0$. This leaves room, when
the temperature is sufficiently low, to use just the local term (of second order in
derivatives) as the main correction to the PFA. But of course, the nonlocal
term will always break down for a higher temperature, whose value will
depend on the actual shape of the surface involved. We stress once more that this non-analytic
behavior is a consequence of the Neumann boundary conditions, and may not be present
for imperfect boundary conditions, as those considered in Ref.\cite{bimonteT}.

In the course of our derivations we have obtained integral expressions for
the momentum space kernels which determine the quadratic contribution to the free energy
for small departure from the planar case. 
Those kernels are well defined in any number of spatial dimensions and
temperatures, and agree with the known results for $d=3$ and $T=0$ \cite{Emig}. 
They can be used to extract the NTLO terms, be they local or nonlocal. 
Although for the static cases we have considered in this article they are
only needed for time independent configurations, we also present the
expressions for the kernels at non-zero frequencies.

\section*{Appendix A: Expansion to order ${\mathcal O}(\eta^2)$, Neumann case.}\label{sec:appendixA}

We present here the main steps and intermediate results corresponding to
the calculation of the free energy up to the second order in the function
$\eta$, which measures the departure from the planar case. We assume that
$\psi(x_\parallel) = a + \eta(x_\parallel)$, with $a$ equaling the average
of $\psi$.
Namely, we want to construct the terms in
\begin{equation}\label{eq:fbn1}
	\Gamma_\beta \;=\; \Gamma^{(0)}_\beta \,+\,  \Gamma^{(1)}_\beta \,+\,
	\Gamma^{(2)}_\beta \,+\,\ldots
\end{equation}
where the index denotes the order in $\eta$. The term of order $1$
vanishes, and, in terms of the expanded matrix elements of ${\mathbb U}$,
we may write the more explicit expressions:
\begin{eqnarray}\label{eq:fbnn}
	\Gamma_\beta^{(0)} &=& \frac{1}{2\beta} \, {\rm Tr} \big[\log {\mathbb
	U}^{(0)}\big]\nonumber\\
	\Gamma_\beta^{(2)} &=&  \Gamma_\beta^{(2,1)} \,+\,\Gamma_\beta^{(2,2)} \;,
\end{eqnarray}
where:
\begin{eqnarray}\label{eq:fbn2}
	\Gamma_\beta^{(2,1)} &=& \frac{1}{2\beta} \, {\rm Tr} \Big[\big({\mathbb
U}^{(0)}\big)^{-1}{\mathbb U}^{(2)}\Big] \nonumber\\
	\Gamma_\beta^{(2,2)} &=& - \frac{1}{4\beta}\, {\rm Tr} 
\Big[\big({\mathbb U}^{(0)}\big)^{-1}{\mathbb U}^{(1)}
\big({\mathbb U}^{(0)}\big)^{-1}{\mathbb U}^{(1)}\Big] \;.
\end{eqnarray}
In order to simplify,  and at the same time render the expressions more
compact, we shall keep the $0$ component of momenta to appear below
continuous, as if they corresponded to zero temperature. In order to obtain
the proper finite temperature expressions one should just replace integrals over the $0$ component of
the momenta by sums over Matsubara frequencies. Namely,
\begin{equation}
	\int \frac{dk_0}{2\pi}\ldots  A(k_0, \ldots) \to \frac{1}{\beta}
	\sum_{n=-\infty}^{+\infty} \; A(\omega_n,\ldots) \;.
\end{equation}
Let us consider the explicit form of $\big({\mathbb U}^{(0)})$ and its
inverse, since both of them are required to construct the terms
contributing to $\Gamma_\beta$ above. We first note that the zero order term is
given by:
\begin{equation}
	{\mathbb U}^{(0)}(x_\parallel,x'_\parallel) \;=\; 
	\left( 
		\begin{array}{cc}
			\partial_d\partial'_d \Delta(x-x')|_{x_d,x'_d\to 0} &
\partial_d\partial'_d\Delta(x-x')|_{x_d\to 0,x'_d\to a}\\
\partial_d\partial'_d\Delta(x-x')|_{x_d\to a,x'_d\to 0}  & 
\partial_d\partial'_d\Delta(x-x')|_{x_d\to a,x'_d\to a}
		\end{array}
	\right) \;,
\end{equation}
which, because of its independence of $\eta$, can be conveniently Fourier
transformed in the parallel coordinates:
\begin{equation}
	{\mathbb U}^{(0)}(x_\parallel,x'_\parallel) \;=\; 
	{\mathbb U}^{(0)}(x_\parallel - x'_\parallel) \;=\; \int
	\frac{d^dk_\parallel}{(2\pi)^d} \, e^{i k_\parallel \cdot
	(x_\parallel - x'_\parallel)} \; \widetilde{\mathbb
	U}^{(0)}(k_\parallel) \;, 
\end{equation}
where
\begin{equation}
	\widetilde{\mathbb U}^{(0)}(k_\parallel) \;=\; -
	\frac{|k_\parallel|}{2}
	\left( 
		\begin{array}{cc}
			1 &  e^{- |k_\parallel| a} \\
		        e^{- |k_\parallel| a}   & 1
		\end{array}
	\right) \;.
\end{equation}

Thus:
\begin{equation}
	\big({\mathbb U}^{(0)}\big)^{-1}(x_\parallel - x'_\parallel) =
\int \frac{d^dk_\parallel}{(2\pi)^d} \, e^{i k_\parallel \cdot (x_\parallel
- x'_\parallel)} \frac{(-2)}{|k_\parallel| (1 - e^{-2 |k_\parallel| a})} 
		\left( \begin{array}{cc}
			1 &  - e^{- |k_\parallel| a} \\
		        - e^{- |k_\parallel| a}   & 1
		\end{array} \right) \;.
\end{equation}

Regarding ${\mathbb U}^{(1)}$, we see that 
\begin{equation}
{\mathbb U}^{(1)}_{RR}= {\mathbb U}^{(1)}_{LL}= 0 \;,
\end{equation}
and
\begin{eqnarray}
	{\mathbb U}^{(1)}_{LR}(x_\parallel,x'_\parallel) &=& \frac{1}{2} 
\int \frac{d^dk_\parallel}{(2\pi)^d} \, e^{i k_\parallel \cdot (x_\parallel
- x'_\parallel)}
\big[ |k_\parallel|^2 \eta(x_\parallel)  + i  k^\alpha \partial_\alpha
\eta(x_\parallel)\big]  e^{- |k_\parallel| a}\nonumber\\
&=& {\mathbb U}^{(1)}_{RL}(x'_\parallel,x_\parallel) \;.
\end{eqnarray}
Finally, to the second order, ${\mathbb U}^{(2)}_{LL}= 0$,  and:
\begin{equation}
	{\mathbb U}^{(2)}_{RR}(x_\parallel,x'_\parallel) = \frac{1}{2} 
\int \frac{d^dk_\parallel}{(2\pi)^d} \, e^{i k_\parallel \cdot (x_\parallel
- x'_\parallel)}
\big( |k_\parallel|^3 + \frac{k^\alpha k^\beta}{|k_\parallel|}
\partial_\alpha \partial'_\beta \big) \;\eta(x_\parallel) \,
\eta(x'_\parallel) \;.
\end{equation}
Regarding ${\mathbb U}^{(2)}_{LR}$ and ${\mathbb U}^{(2)}_{LR}$, they are non-vanishing, 
but it may be seen that they do not contribute to the second order term. 

Thus,
\begin{eqnarray}
	\Gamma_\beta^{(2,1)} &=& \frac{1}{2\beta} \, 
	{\rm Tr} \Big[\big({\mathbb U}_{RR}^{(0)}\big)^{-1}{\mathbb
	U}_{RR}^{(2)}\Big] \nonumber\\
	&=& \frac{1}{2\beta} \, \int d^dx_\parallel d^dx'_\parallel \, 
	\big({\mathbb
		U}_{RR}^{(0)}\big)^{-1}(x_\parallel,x'_\parallel) \, 
	{\mathbb U}_{RR}^{(2)}(x'_\parallel,x_\parallel) \;,
\end{eqnarray}
and (using the properties of the matrix elements under the exchange of
arguments):
\begin{eqnarray}
	\Gamma_\beta^{(2,2)} &=& - \frac{1}{2\beta} \, 
	{\rm Tr} \Big[\big({\mathbb U}_{LL}^{(0)}\big)^{-1}{\mathbb
	U}_{LR}^{(1)}\big({\mathbb U}_{RR}^{(0)}\big)^{-1}{\mathbb
	U}_{RL}^{(1)}\Big] 
	\nonumber\\
	&-& \frac{1}{2\beta} \, 
	{\rm Tr} \Big[\big({\mathbb U}_{RL}^{(0)}\big)^{-1}{\mathbb
	U}_{LR}^{(1)}\big({\mathbb U}_{RL}^{(0)}\big)^{-1}{\mathbb
	U}_{LR}^{(1)}\Big] \;.
\end{eqnarray}

In Fourier space, after some algebra, one then finds:
\begin{equation}
	\Gamma_\beta^{(2)} \;=\; \frac{1}{2} \, \int
	\frac{d^dk_\parallel}{(2\pi)^d} \, g^{(2)}(k_\parallel) \,
	\big| \eta(k_\parallel) \big|^2 \;,
\end{equation}
with:
\begin{eqnarray}
g^{(2)}(k_\parallel)  = - 2\, \int \frac{d^dp_\parallel}{(2\pi)^d}
	\frac{\big[p_\parallel \cdot (p_\parallel + k_\parallel)\big]^2}{|p_\parallel| \, 
	|p_\parallel + k_\parallel|} \, 
\frac{1}{1 - e^{- 2a |p_\parallel|}}  \frac{1}{e^{2a |p_\parallel + k_\parallel|} - 1 } \;.
\label{g2ddim}
\end{eqnarray}

\section*{Appendix B: Exact expressions for the form factors in $3+1$ dimensions.}\label{sec:appendixB}

In this Appendix we present exact expressions and series expansions for the form factors $f^{(2)}(k_\parallel)$ and $g^{(2)}(k_\parallel)$ at zero temperature
and $d=3$. These expressions have been previously obtained in Ref.\cite{Emig} (see also Ref.\cite{Miri}). We will use the dimensionless
quantity $x=a\vert k_\parallel\vert$.

For Dirichlet boundary conditions the form factor $f^{(2)}$ reads
\begin{eqnarray}
f^{(2)}(k_\parallel)  = - 2\, \int \frac{d^3p_\parallel}{(2\pi)^3}
\frac{|p_\parallel| \, 
	|p_\parallel + k_\parallel|}{(1 - e^{- 2a |p_\parallel|})(e^{2a |p_\parallel + k_\parallel|} - 1)}\, .
\end{eqnarray}
An explicit evaluation gives \cite{Emig}:
\begin{eqnarray}
a^5 f^{(2)}(x) &=&
-\frac{x^3 \text{Li}_2\left(e^{-2 x}\right)}{48 \pi ^2}-\frac{x^2
   \text{Li}_3\left(e^{-2 x}\right)}{24 \pi ^2}-\frac{x
   \text{Li}_4\left(e^{-2 x}\right)}{16 \pi
   ^2}-\frac{\text{Li}_5\left(e^{-2 x}\right)}{16 \pi ^2}-\frac{\pi ^2
   \text{Li}_2\left(1-e^{-2 x}\right)}{240
   x}\nonumber\\
   &+& \frac{\frac{\pi ^6}{945}-\text{Li}_6\left(e^{-2 x}\right)}{32 \pi
   ^2 x}+\frac{x^4 \log \left(1-e^{-2 x}\right)}{120 \pi ^2}-\frac{\pi ^2
   x}{240},
\end{eqnarray}
where  $\rm{Li}_n$ denote Polylogarithm functions. This result can be expanded in powers of $x$ as follows:
\begin{eqnarray}
 a^5 f^{(2)}(x) &\approx&
-\frac{\pi ^2}{120}-\frac{\pi ^2 x^2}{1080}+\frac{\left(45+\pi ^4\right)
   x^4}{27000 \pi ^2}-\frac{x^5}{720 \pi ^2}+\frac{\left(315-2\pi^4\right) x^6}{793800 \pi ^2}
   \nonumber\\
   && +\frac{\left(\pi ^4-105\right)
   x^8}{5103000 \pi ^2}+\frac{\left(\frac{8}{165}-\frac{8 \pi
   ^4}{16335}\right) x^{10}}{30240 \pi
   ^2}+O\left(x^{12}\right)
   \label{seriesgtm}
   \end{eqnarray}
which shows the presence of a non-analytic term proportional to $x^5$, which is independent of $a$.

For Neumann boundary conditions  the form factor $g^{(2)}$ is given in Eq.(\ref{g2ddim}) with $d=3$. Evaluating explicitly 
this integral it is possible to show that \cite{Emig}
\begin{eqnarray}
&& a^5 g^{(2)}(x) =  \frac{1}{24} \left(\frac{x^2}{2 \pi ^2}+1\right) x \text{Li}_2\left(e^{-2
   x}\right)+\left(\frac{1}{16}-\frac{x^2}{48 \pi ^2}\right)
   \text{Li}_3\left(e^{-2 x}\right)-\frac{5 x \text{Li}_4\left(e^{-2
   x}\right)}{32 \pi ^2}-\frac{7 \text{Li}_5\left(e^{-2 x}\right)}{32 \pi
   ^2} \nonumber\\
   &&
   -\frac{\pi ^2 \text{Li}_2\left(1-e^{-2 x}\right)}{240 x}+\frac{\pi ^2
   \text{Li}_4\left(e^{-2 x}\right)-\frac{7 \text{Li}_6\left(e^{-2
   x}\right)}{2}-\frac{\pi ^6}{135}}{32 \pi ^2 x}+\frac{x^4 \log \left(1-e^{-2
   x}\right)}{120 \pi ^2}-\frac{\pi ^2 x}{720}\, .
\end{eqnarray}
Expanding this result for $x\ll 1$ we obtain
\begin{eqnarray}
&& a^5 g^{(2)}(x)   \approx 
-\frac{\pi ^2}{120}+\frac{\left(30-\pi ^2\right)
   x^2}{1080}-\frac{x^3}{64}+\frac{\left(1095+50 \pi ^2+\pi ^4\right) x^4}{27000
   \pi ^2}-\frac{11 x^5}{720 \pi ^2}+\frac{\left(2205-42 \pi ^2-2 \pi ^4\right)
   x^6}{793800 \pi ^2}+ \nonumber \\
   &&
   \frac{\left(100 \pi ^2+7 \pi ^4-3045\right) x^8}{35721000
   \pi ^2}+4 \left(\frac{47}{38102400 \pi ^2}-\frac{1}{23328000}-\frac{\pi
   ^2}{246985200}-\frac{\frac{1}{1403325}+\frac{\pi ^2}{4677750}}{128 \pi
   ^2}\right) x^{10}+ O(x^{12})\, .
   \label{seriesgte}
\end{eqnarray}
We see that the Neumann form factor has non-analytic terms proportional to $x^3$ and $x^5$.

\section*{Acknowledgements}
This work was supported by ANPCyT, CONICET, UBA and UNCuyo.


\begin{thebibliography}{bib}
\bibitem{footnote} In the case of two flat and parallel mirrors, these excitations correspond to  modes that are independent of $x_3$,
but depend on $x_1$ and $x_2$. Given that the transverse dimensions of the mirrors are very large, there is no gap for the excitation of these modes
when one of the mirrors is in motion.
\end{thebibliography}
\end{document}